\documentclass[12pt]{iopart}
\usepackage{iopams}
\usepackage{subcaption}
\usepackage{graphicx}
\usepackage{xcolor}

\newcommand{\LT}[1]{\widetilde{#1}}
\newcommand{\FT}[1]{\widehat{#1}}
\newcommand{\FPT}[1]{\tau_{#1}}
\newcommand{\wpr}{\widehat{\wp}_R}
\newcommand{\wpi}{\widehat{\wp}_I}
\newcommand{\sgn}{\mathrm{sgn}}

\newcommand{\coloneq}{:=}
\begin{document}

\title[First-passage statistics of random walks]{First-passage statistics of random walks: a general approach via Riemann-Hilbert problems}
\author{Mattia Radice$^1$, Giampaolo Cristadoro$^3$}

\address{$^1$Dipartimento di Fisica e Astronomia, Università di Bologna, 40127 Bologna, Italy}
\address{$^2$INFN Sezione di Bologna, 40127 Bologna, Italy}
\address{$^3$Dipartimento di Matematica e Applicazioni, Università degli Studi di Milano - Bicocca, 20126 Milan, Italy}
\ead{\mailto{mattia.radice@unibo.it}}
\vspace{10pt}

\begin{abstract}
We study first-passage statistics for one-dimensional random walks $S_n$ with independent and identically distributed jumps starting from the origin. We focus on the joint distribution of the first-passage time $\FPT{b}$ and first-passage position $S_{\FPT{b}}$ beyond a threshold $b\geq0$, as well as the distribution of $S_n$ for the walks that do not cross $b$ up to step $n$. By solving suitable Riemann-Hilbert problems, we are able to obtain exact and semi-explicit general formulae for the quantities of interest.  Notably, such formulae are written solely in terms of the characteristic function of the jumps. In contrast with previous results, our approach is universally valid, applicable to both continuous and discrete, symmetric and asymmetric jump distributions.  We complement our theoretical findings with explicit examples.
\end{abstract}

%
\vspace{2pc}
%
%
%
%

\section{Introduction}
Random walks represent a foundational theoretical tool for modelling complex phenomena. Their importance is demonstrated by their presence in many areas of natural science, as they emerge in the description of fluctuating quantities in physics \cite{vanKam}, especially in statistical mechanics \cite{Set}, as well as in related fields such as chemistry \cite{vanKam}, finance \cite{Set,Gar} and biology \cite{AltMorWal-2020}.

In their simplest formulation, random walks are described by a sum of independent and identically distributed (iid) random variables. Many interesting questions are related to the amount of fluctuations that such a sum undergoes, which is the object of study of \emph{fluctuation theory} \cite{Por-1963}. A class of problems much studied in this area is represented by the so-called \emph{first-passage problems}, which study the occurrence of a given condition, such as the crossing of a threshold, for the first time. First-passage problems are of significant importance to a considerable number of applications \cite{Fell-II,Red,Metzler2014} and especially relevant in Extreme Value Statistics (EVS), where they are associated with the analysis of extreme events, such as maxima and records.

To fix the ideas, let us consider a one-dimensional random walk $S_n$ starting from the origin. Suppose that $b\geq0$ represents a threshold or the position of a barrier. Typical quantities of interest in fluctuation theory are:
\begin{enumerate}
	\item the \emph{first-passage time} $\FPT{b}$, which is a random variable corresponding to the number of steps required to exceed $b$ for the first time. The first-passage time distribution has applications in several fields, ranging from physics to economics \cite{Gar,Red,Metzler2014};
	\item the value of the sum $S_n$ upon first passage, namely, the \emph{first-passage position} $S_{\FPT{b}}$. Note that the quantity $L_b\coloneq S_{\FPT{b}}-b$, which is called \emph{overshoot} or \emph{leap-over} \cite{Kor-Lom-al}, is relevant for search processes \cite{PalCheMet-2014}, in renewal theory \cite{Barkai-2003}, and has been recently considered in the context of resetting \cite{RADCri-2024}. In the case $b=0$, the first-passage position corresponds to the first positive value of the random walk, which is known in the mathematical literature as \emph{first ladder height}. The distribution of the first ladder height is a major object of investigation in fluctuation theory \cite{Fell-II,Sin-1957,Ray-1958,Rog-1964,DonGre-1993,DonGre-1995}, even beyond the case of iid variables \cite{BiaCriPoz-2025}. The statistical physics community has emphasized its importance in the study of order statistics \cite{MajMouSch-2013,MajMouSch-2014} and records' increments \cite{GodMajSch-2016,GodMajSch-2017}. For a very recent work, see \cite{God-Luc-2025} and references therein.
	\item the \emph{survival probability}, namely, the probability that $b$ is never exceeded up to time $n$. The survival probability $q_b(n)\coloneq\mathbb{P}(S_1\leq b; S_2\leq b; \dots,S_n\leq b)$ corresponds to $\mathbb{P}(M_n\leq b)$, where $M_n\coloneq\max(0,S_1,\dots,S_n)$. It is thus a fundamental quantity in EVS \cite{Fell-II,Maj-2005,Maj-2010,AOPR} and finds an impressive number of applications involving the analysis of extreme events and records \cite{Maj-2010,MajComZif-2005,ComMaj-2005,ArtCriDeg-2014,MajMouSch-2017}. Furthermore, it is crucial to study persistence and critical phenomena for equilibrium and non-equilibrium systems \cite{MajSire-1996,OerCorBra-1997,MajRosZoi-2010,BraMajSch-2013}. 
\end{enumerate}
These quantities have long been of interest to both mathematicians and physicists.

In the mathematical literature, contributions in this direction were given by Hopf \cite{Hop}, Pollaczek \cite{Pollaczek-1952}, Sparre-Andersen \cite{Sparre}, Baxter \cite{Bax-1958} and Spitzer \cite{Spi-1956,Spi-1957,Spi-1960}, to name but a few. In particular, Spitzer made use of a method, which is similar to the Wiener-Hopf technique and had been previously used by Ray \cite{Ray-1958}, to prove a remarkable identity concerning the joint distribution of the first-passage time and position in the case $b=0$ (\emph{Baxter-Spitzer identity}) \cite{Spi-1960}. Since then, the Wiener-Hopf method has been widely used to derive various fluctuation identities \cite{Fell-II}, proving particularly useful in mathematical finance applications \cite{GreFusAbr-2010,FusGerMar-2016}.

In the physics literature, some of these results have proved especially influential. The statistical physics community has made extensive use of the so-called Pollaczek-Spitzer \cite{Maj-2010,Pollaczek-1952,Spi-1957} and Hopf-Ivanov \cite{MajMouSch-2014,Hop,Iva-1994} identities. The advantage of these formulae over others is that they express the transforms of the desired distributions only in terms of the common characteristic function of the jumps $X_i$. This allows one to obtain precise asymptotic results, even when explicit formulae are not available, see e.g. \cite{ComMaj-2005,MajMouSch-2017}. However, their validity is limited to the case of symmetric and continuous jump distributions. The purpose of this paper is to provide a generalization of these formulae valid also for asymmetric jump distributions, as well as their version for the case of discrete jumps. Although some possible generalisations have been already considered \cite{Maj-2010,LeDWie-2009,MajSchWer-2012,MouMajSch-2018,MouMajSch-2020,BurMaj-2025-arxiv}, as far as we know there is no universal framework in the physics literature, as the one we propose here.

The method we will use is closely related to the Wiener-Hopf technique, and is based on the solution of a Riemann-Hilbert problem. It must be noted that the Wiener-Hopf and the Riemann-Hilbert methods share many similarities, so much so that one is often associated with the other. However, it is important to emphasize that the Wiener-Hopf technique requires the factorization of a function that is analytic on a strip in the complex plane \cite{Nob}, while in a Riemann-Hilbert problem one only requires a continuity condition (Hölder condition) on a curve \cite{Gak}. Since in our case the object to be factorised is an expression containing the characteristic function of the jumps, our choice is also suitable for cases in which the analyticity of the characteristic function is not guaranteed. We refer the reader to \cite{Kis-2015} for a detailed discussion on the differences between the methods, and to \cite{KisAbrMisRog-2021} for a recent review.

The paper is organised as follows: in \sref{s:Model} we describe the model and present the main formulae; in  \sref{s:Examples} we apply such formulae to a couple of examples that admit closed-form results; \sref{s:Applications} is dedicated to cases that, while not permitting closed-form expressions, enable the derivation of exact asymptotic estimates; the mathematical derivation of the formulae, which result from the solution of suitable Riemann-Hilbert problems, is presented in \sref{s:derivation}; finally, in \sref{s:Concl} we draw our conclusions.

\section{Definition of the model and main results}\label{s:Model}
Consider a one-dimensional discrete-time random walk starting from the origin
\begin{equation}\label{eq:RW_dyn}
	S_{n}=S_{n-1}+X_{n},\quad n\ge1,\quad S_0=0,
\end{equation}
where $X_i$ are iid real-valued random variables. Denote with $\FT{\wp}(k)$ the common characteristic function of the jumps $X_i$, viz. $\FT{\wp}(k)\coloneq\mathbb{E}(\rme^{\rmi kX_1}) $.
We set $b\geq0$ and call $\FPT{b}$ the first-passage time in $(b,\infty)$, namely
\begin{equation}
	\FPT{b}\coloneq\min\left\{n>0\,|\,S_n>b\right\}.
\end{equation}
Furthermore, we introduce the distribution functions
\numparts
\begin{eqnarray}
	F_b(x;n)&\coloneq\mathbb{P}(S_n\leq x; \FPT{b}=n)\\
	Q_b(x;n)&\coloneq\mathbb{P}(S_n\leq x; \FPT{b}>n),
\end{eqnarray}
\endnumparts
so that $\int_I dF_b(x;n)$ is the probability that the first-passage position $S_{\FPT{b}}$ lies in $I$, and $\int_I dQ_b(x;n)$ is the probability that $S_{n}\in I$ without ever leaving $(-\infty,b]$. These are the central quantities for determining the fundamental objects in fluctuation theory mentioned above. For example, the first-passage time in $(b,\infty)$ can be obtained as $\tau_b=\int_{b}^{\infty}dF_b(x;n)$, and the survival probability in $(-\infty,b]$ as $q_b(n)=\int_{-\infty}^{b}dQ_b(x;n)$.

Define the transforms
\numparts
\begin{eqnarray}
	\mathcal{F}_b(k,\zeta)&\coloneq\sum_{n=1}^{\infty}\zeta^n\int_{b}^{\infty}\rme^{\rmi kx}dF_b(x;n)\label{eq:Fb_def}\\
	\mathcal{Q}_b(k,\zeta)&\coloneq\sum_{n=0}^{\infty}\zeta^n\int_{-\infty}^{b}\rme^{\rmi kx}dQ_b(x;n),\label{eq:Qb_def}
\end{eqnarray}
\endnumparts
where $0<\zeta<1$. Our approach starts from a well-known relation between $\mathcal{F}_b(k,\zeta)$ and $\mathcal{Q}_b(k,\zeta)$, see e.g. \cite{Fell-II}, which reads
\begin{equation}\label{eq:fund_eq_trans0}
	1-\mathcal{F}_b(k,\zeta)=\left[1-\zeta\FT{\wp}(k)\right]\mathcal{Q}_b(k,\zeta).
\end{equation}
Notably, in this equation the two transforms are related by means of the characteristic function $\FT{\wp}(k)$, which depends only on the jump distribution. Our problem consists in obtaining $\mathcal{F}_b(k,\zeta)$ and $\mathcal{Q}_b(k,\zeta)$ from \eref{eq:fund_eq_trans0}. By properly interpreting this problem as a non-homogeneous Riemann–Hilbert problem, we obtain exact general expressions  for the transforms $\mathcal{F}_b(k,\zeta)$ and $\mathcal{Q}_b(k,\zeta)$, which are written  in terms of $\FT{\wp}(k)$ only. In this section we list such formulae,  postponing the detail of their derivation to \sref{s:derivation}.

\subsection{Main formulae}
The results use the notion of \emph{Cauchy principal value} of a singular integral, see \ref{s:App_formule} for more details. Contrary to the common practice of using a special symbol, we will follow Gakhov \cite{Gak} and denote the principal value as a regular integral, with the convention that each singular integral appearing in the text has to be interpreted as a principal value. For reasons that we will clarify later, we provide different sets of formulae for continuous and discrete jumps.

\paragraph{Continuous jumps.}
The transforms $\mathcal{F}_b(k,\zeta)$ and $\mathcal{Q}_b(k,\zeta)$ are given by:
\numparts
\begin{eqnarray}
	\mathcal{F}_b(k,\zeta)&=1-\sqrt{1-\zeta\FT{\wp}(k)}\mathcal{H}_b(k,\zeta)\rme^{-\rmi\Omega(k,\zeta)}\label{eq:Fb_cont}\\
	\mathcal{Q}_b(k,\zeta)&=\frac{\mathcal{H}_b(k,\zeta)}{\sqrt{1-\zeta\FT{\wp}(k)}}\rme^{-\rmi\Omega(k,\zeta)},\label{eq:Qb_cont}
\end{eqnarray}
\endnumparts
where the phase function $\Omega(k,\zeta)$ is defined as
\begin{equation}\label{eq:Omega}
	\Omega(k,\zeta)=\frac{1}{2\pi}\int_{-\infty}^{\infty}\frac{\ln\left[1-\zeta\FT{\wp}(t)\right]}{t-k}dt.
\end{equation}
To give the expression of the factor $\mathcal{H}_b(k,\zeta)$, let us introduce the auxiliary function
\begin{equation}\label{eq:Psi_0^+_in_main_formulae}
	\Psi_0^+(z,\zeta)=\exp\left\{\frac1{2\pi\rmi}\int_{-\infty}^{\infty}\frac{\ln[1-\zeta\FT{\wp}(t)]}{t-z}dt\right\},\quad\Im(z)>0,
\end{equation}
where $z$ is a complex number with positive imaginary part. Then, $\mathcal{H}_b(k,\zeta)$ is given by
\begin{equation}\label{eq:Hb}
	\mathcal{H}_b(k,\zeta)=\frac{1}{2\pi \rmi}\int_{\mathcal{B}}\frac{\rme^{ub}}{u}\frac{du}{\Psi_0^+(k+\rmi u,\zeta)},
\end{equation}
where $\mathcal{B}$ is a vertical line along the imaginary axis, deformed in such a way that $u=0$ remains on the left. For $b=0$ one gets $\mathcal{H}_b(k,\zeta)=1$, see the proof in \sref{s:derivation}, and thus we recover the same formulae for the case $b=0$ that were obtained in \cite{Kla-Sok}.

\paragraph{Discrete jumps.}
The expressions of the transforms $\mathcal{F}_b(k,\zeta)$ and $\mathcal{Q}_b(k,\zeta)$ are:
\numparts
\begin{eqnarray}
	\mathcal{F}_b(k,\zeta)&=1-\sqrt{1-\zeta\FT{\wp}(k)}B(\zeta)H_b(k,\zeta)\rme^{-\rmi \omega(k,\zeta)}\label{eq:Fb_disc}\\
	\mathcal{Q}_b(k,\zeta)&=\frac{B(\zeta)H_b(k,\zeta)}{\sqrt{1-\zeta\FT{\wp}(k)}}\rme^{-\rmi\omega(k,\zeta)},\label{eq:Qb_disc}
\end{eqnarray}
\endnumparts
where the phase $\omega(k,\zeta)$ is now defined as
\begin{equation}
	\omega(k,\zeta)= \frac{1}{4\pi}\int_{-\pi}^{\pi}\ln[1-\zeta\FT{\wp}(\theta)]\cot\left(\frac{\theta-k}2\right)d\theta.\label{eq:omega}
\end{equation}
The factors $B(\zeta)$ and $H_b(k,\zeta)$ are defined as follows: let $t$ be a complex number on the unit circle, with $|\arg(t)|<\pi$, and define
\begin{equation}\label{eq:char_fun_*}
	\FT{\wp}_*(t)\coloneq\FT{\wp}(-\rmi\ln t)=\mathbb{E}\left(t^{X_1}\right).
\end{equation}
Furthermore, let us introduce the complex variable $w=\rho\rme^{\rmi\theta}$ and the auxiliary function
\begin{equation}\label{eq:phi_0^+_in_main_formulae}
	\phi_0^+(w,\zeta)=\exp\left\{\frac1{2\pi\rmi}\int_{\mathcal{C}}\frac{\ln[1-\zeta\FT{\wp}_*(t)]}{t-w}dt\right\},\quad|w|<1,
\end{equation}
where $\mathcal{C}$ is the unit circle. Then, $B(\zeta)$ is given by
\begin{equation}
	B(\zeta)=\frac{1}{\sqrt{\phi_0^+(0,\zeta)}}=\exp\left\{-\frac{1}{4\pi}\int_{-\pi}^{\pi}\ln[1-\zeta\FT{\wp}(\theta)]d\theta\right\},\label{eq:B}
\end{equation}
and $H_b(k,\zeta)$ is
\begin{equation}\label{eq:Hb_disc}
	H_b(k,\zeta)=\phi_0^+(0,\zeta)\mathrm{Res}\left[\frac{w^{-b-1}}{\phi_0^+(w,\zeta)}\frac{\rme^{\rmi k(b+1)}}{\rme^{\rmi k}-w},w=0\right],
\end{equation}
where $\mathrm{Res}[f(z),z=z_0]$ denotes the residue of $f(z)$ at the point $z_0$. Note that $H_0(k,\zeta)=1$, see \sref{s:derivation}, thus formula \eref{eq:Qb_disc} agrees with and generalizes a result of \cite{MouMajSch-2020}, where an exact expression for $\mathcal{Q}_b(k,\zeta)$ was obtained in the case of symmetric jumps and $b=0$.

\section{Explicit examples: random walks admitting closed-form results}\label{s:Examples}
It is apparent from the general expressions of the previous section that extracting closed-form result requires the ability to perform exact calculations, which is usually very difficult. However, as we shall see in \sref{s:derivation}, the calculations become trivial, or can be avoided altogether, when one can factorize the function $1-\zeta\FT{\wp}(k)$ as
\begin{equation}\label{eq:factorization}
	1-\zeta\FT{\wp}(k)=\frac{\psi^+(k,\zeta)}{\psi^-(k,\zeta)},
\end{equation}
where $\psi^\pm(k,\zeta)$ can be extended to suitable bounded analytic function in the upper and lower half-planes, respectively. Such a factorization is possible, for instance, when $\FT{\wp}(k)$ is a polynomial.
 A well-known case corresponds to the symmetric Erlang distribution, which has been studied deeply by the statistical physics community \cite{God-Luc-2025,LucFunNie-1991,BatMajSch-2020}. Another solvable case corresponds to the sum of symmetric Laplace distributions \cite{God-Luc-2025}. 

In the following examples, we present situations for both continuous and discrete jump distributions in which, despite the presence of asymmetry in the jumps, the factorization is easy to find, allowing for a thorough analysis.

\subsection{Explicit example: Skewed Laplace distribution}
Suppose that the distribution of the jumps $X_i$ is characterized by the following density, which we introduced in \cite{RADCriTha-2025}:
\begin{equation}
	\wp(\xi)=\frac{1}{\sqrt{\mu^2+4\delta^2}}\exp\left[-\frac1{2\delta^2}\left(\sqrt{\mu^2+4\delta^2}|\xi|-\mu\xi\right)\right].
\end{equation}
Here $-\infty<\mu<\infty$ corresponds to the mean jump, viz. $\int_{-\infty}^{\infty}\xi\wp(\xi)d\xi=\mu$, while $\delta>0$ is a length scale related to the variance $\sigma^2$ and $\mu$ by $\sigma^2=2\delta^2+\mu^2$. Note that for $\mu=0$ we recover a symmetric Laplace random variable, with $\wp(\xi)=\exp(-|\xi|/\delta)/2\delta$. One can show that
\begin{equation}
	\FT{\wp}(k)=\frac{1}{1-\rmi \mu k+\delta^2k^2},
\end{equation}
which corresponds to the characteristic function of the sum of two exponential random variables $E^{\pm}$, which are concentrated on $(0,\infty)$ and $(-\infty,0)$, respectively, and whose means are $\mu^+=\mu/2+\sqrt{\delta^2+\mu^2/4}$ and $\mu^-=\mu/2-\sqrt{\delta^2+\mu^2/4}$.

We observe that
\begin{equation}
	1-\zeta\FT{\wp}(k)=\frac{1-\zeta-\rmi \mu k+\delta^2k^2}{1-\rmi \mu k+\delta^2k^2},
\end{equation}
which can be factorized as in \eref{eq:factorization}, with
\begin{eqnarray}
	\psi^+(k,\zeta)&=\frac{r_-(\zeta)-\rmi \delta k}{r_-(0)-\rmi \delta k}\\
	\psi^-(k,\zeta)&=\frac{r_+(0)+\rmi \delta k}{r_+(\zeta)+\rmi \delta k},
\end{eqnarray}
where $r_-(\zeta)$ and $r_+(\zeta)$ are positive numbers defined by
\begin{eqnarray}
	r_-(\zeta)&=\frac1{2\delta}\left[\sqrt{\mu^2+4\delta^2(1-\zeta)}-\mu\right]\label{eq:r_l}\\
	r_+(\zeta)&=\frac1{2\delta}\left[\sqrt{\mu^2+4\delta^2(1-\zeta)}+\mu\right].\label{eq:r_u}
\end{eqnarray}
We note that $\psi^\pm(z,\zeta)$ are analytic and bounded for $\Im(z)>0$ and $\Im(z)<0$, respectively. Furthermore, $\psi^+(z,\zeta)\to1$ as $|z|\to\infty$ in the upper-half plane, and similarly, $\psi^-(z,\zeta)\to1$ as $|z|\to\infty$ in the lower-half plane. We will prove in \sref{s:derivation} that this is sufficient to conclude that $\Psi_0^+(z,\zeta)=\psi^+(z,\zeta)$. Here instead we demonstrate this equality by direct calculation using \eref{eq:Psi_0^+_in_main_formulae}. Note that due to the factorization, we can write
\begin{eqnarray}
	\lim_{R\to\infty}\int_{-R}^{R}\frac{\ln[1-\zeta\FT{\wp}(t)]}{t-z}dt=&\lim_{R\to\infty}\int_{-R}^{R}\frac{\ln\psi^+(t,\zeta)}{t-z}dt\nonumber\\
	&-\lim_{R\to\infty}\int_{-R}^{R}\frac{\ln\psi^-(t,\zeta)}{t-z}dt,
\end{eqnarray}
where $\Im(z)>0$. To compute the first integral at the rhs, we consider a contour in the complex plane consisting of the segment $(-R,R)$ and the semicircle $z=R\rme^{\rmi\theta}$, with $0<\theta<\pi$. Recalling the properties of $\psi^+(z,\zeta)$, an easy application of the Cauchy theorem yields
\begin{equation}
	\lim_{R\to\infty}\int_{-R}^{R}\frac{\ln\psi^+(t,\zeta)}{t-z}dt=2\pi\rmi\ln\psi^+(z,\zeta).
\end{equation}
To evaluate the second integral, we consider a similar contour, consisting of the segment $(-R,R)$ and the semicircle $z=R\rme^{\rmi\theta}$, this time with $-\pi<\theta<0$. Again, due to the properties of $\psi^-(z,\zeta)$, according to the Cauchy theorem the integral yields a vanishing contribution in the limit $R\to\infty$. Putting all together, we have
\begin{equation}
	\int_{-\infty}^{\infty}\frac{\ln[1-\zeta\FT{\wp}(t)]}{t-z}dt=2\pi\rmi\ln\psi^+(z,\zeta),
\end{equation}
and the equality $\Psi_0^+(z,\zeta)=\psi^+(z,\zeta)$ follows from formula \eref{eq:Psi_0^+_in_main_formulae}.

As a second step, we determine the phase $\Omega(k,\zeta)$. This task can be easily performed by employing formula \eref{eq:App_SokPle+}, which leads to the equality
\begin{equation}
	\rme^{-\rmi\Omega(k,\zeta)}=\frac{\Psi_0^+(k,\zeta)}{\sqrt{1-\zeta\FT{\wp}(k)}}.\label{eq:Skw_Laplace_Omega}
\end{equation}

Finally, we compute $\mathcal{H}_b(k,\zeta)$ from \eref{eq:Hb}:
\begin{equation}
	\mathcal{H}_b(k,\zeta)=\frac{1}{2\pi \rmi}\int_{\mathcal{B}}\frac{\rme^{ub}}{u}\frac{du}{\Psi_0^+(k+\rmi u,\zeta)}=\frac{1}{2\pi \rmi}\int_{\mathcal{B}}\frac{\rme^{ub/\delta}}{u}\frac{u+r_-(0)-\rmi \delta k}{u+r_-(\zeta)-\rmi \delta k}du.
\end{equation}
The integrand has two simple poles, $u_0=0$ and $u_1=\rmi \delta k-r_-(\zeta)$, with non-positive real part. Thus by the residue theorem we get
\begin{eqnarray}
	\mathcal{H}_b(k,\zeta)&=\frac{r_-(0)-\rmi \delta k}{r_-(\zeta)-\rmi \delta k}-\frac{r_-(0)-r_-(\zeta)}{r_-(\zeta)-\rmi \delta k}\rme^{-r_-(\zeta)b/\delta+\rmi kb}\\
	&=\frac1{\Psi_0^+(k,\zeta)}\left[1-\frac{r_-(0)-r_-(\zeta)}{r_-(0)-\rmi \delta k}\rme^{-r_-(\zeta)b/\delta+\rmi kb}\right].
\end{eqnarray}
Therefore, \eref{eq:Fb_cont} and \eref{eq:Qb_cont} finally yield
\numparts
\begin{eqnarray}
	\mathcal{F}_b(k,\zeta)&=\frac{r_-(0)-r_-(\zeta)}{r_-(0)-\rmi \delta k}\rme^{-r_-(\zeta)b/\delta+\rmi kb}\\
	\mathcal{Q}_b(k,\zeta)&=\frac{1}{1-\zeta\FT{\wp}(k)}\left[1-\frac{r_-(0)-r_-(\zeta)}{r_-(0)-\rmi \delta k}\rme^{-r_-(\zeta)b/\delta+\rmi kb}\right].
\end{eqnarray}
\endnumparts
We can now take advantage of these results to derive the interesting quantities of fluctuation theory.

\paragraph{Escape probability.} The transform $\mathcal{F}_b(0,\zeta)$ is the generating function of $f_b(n)\coloneq\mathbb{P}(\FPT{b}=n)$. Therefore, the escape probability from $(-\infty,b]$ is
\begin{equation}
	\mathcal{E}_b\coloneq\sum_{n=1}^{\infty}f_b(n)=\lim_{\zeta\to1}\mathcal{F}_b(0,\zeta).
\end{equation}
A straightforward computation yields
\begin{equation}
	\mathcal{E}_b=\cases{
	1 & if $\mu\geq0$\\
	\left(\frac{\sqrt{\mu^2+4\delta^2}-|\mu|}{\sqrt{\mu^2+4\delta^2}+|\mu|}\right)\rme^{-|\mu| b/\delta^2} & if $\mu<0$
	}
\end{equation}
Thus, if the jumps are biased in the direction of $b$ or unbiased, the random walk eventually escapes with probability one. Otherwise, there is a non-zero fraction of walks that never escape. Remarkably, if we renormalize $\mathcal{F}_b(0,\zeta)$ by $\mathcal{E}_b$, we get
\begin{equation}\label{eq:Skw_Laplace_F_over_E}
	\frac{\mathcal{F}_b(0,\zeta)}{\mathcal{E}_b}=\frac{\sqrt{\mu^2+4\delta^2}-\sqrt{\mu^2+4\delta^2(1-\zeta)}}{\sqrt{\mu^2+4\delta^2}-|\mu|}\rme^{-\left[\sqrt{\mu^2+4\delta^2(1-\zeta)}-|\mu|\right]\case b{2\delta^2}},
\end{equation}
which is independent of the sign of $\mu$. It follows that the first-passage time distribution of the walks biased towards $b$ is the same as the walks biased away from $b$, provided that we condition on the event that the walks actually escape. This phenomenon is called in some contexts \emph{first-passage duality} \cite{KraRed-2018}.

\paragraph{First-passage time.} In principle, by evaluating the derivatives of $\mathcal{F}_b(0,\zeta)$ with respect to $\zeta$ at $\zeta=0$, one can compute $f_b(n)$ for each $n$. In most cases, one is interested in the large-$n$ behaviour of $f_b(n)$, which is related by Tauberian theorems to the behaviour of $\mathcal{F}_b(0,\zeta)$ as $\zeta\to1$. We have seen that $\mathcal{F}_b(0,\zeta)$ converges to a constant, which means that $f_b(n)$ decays faster than $1/n$ in any case. So we consider
\begin{eqnarray}
	\frac{\partial\mathcal{F}_b(0,\zeta)}{\partial\zeta}&=\sum_{n=1}^{\infty}nf_b(n)\zeta^{n-1}\\
	&=-r_-'(\zeta)\rme^{-r_-(\zeta)b/\delta}\left[\frac1{r_-(0)}+\frac b\delta\left(1-\frac{r_-(\zeta)}{r_-(0)}\right)\right].
\end{eqnarray}
If $\mu=0$, we have $r_-(\zeta)=\sqrt{1-\zeta}$ and $r_-'(\zeta)=-1/(2\sqrt{1-\zeta})$. Thus, for $\zeta\to1$ and $b$ fixed, the rhs behaves as
\begin{equation}
	\frac{1+b/\delta}{2\sqrt{1-\zeta}},
\end{equation}
whence it follows
\begin{equation}
	f_b(n)\sim\frac{1+b/\delta}{\sqrt{4\pi n^3}}.
\end{equation}
If $\mu\neq0$ instead, we have $r_-(\zeta)=(\sqrt{\mu^2+4\delta^2(1-\zeta)}-\mu)/2\delta$ and $r_-'(\zeta)=-\delta/[\sqrt{\mu^2+4\delta^2(1-\zeta)}]$, thus the behaviour of the rhs is still not singular. It is useful to rewrite $\zeta=(\case{\mu^2}{4\delta^2}+1)\eta$, so that
\begin{eqnarray}
	\left.\frac{\partial\mathcal{F}_b(0,\zeta)}{\partial\zeta}\right|_{\zeta=\left(\case{\mu^2}{4\delta^2}+1\right)\eta}&=\sum_{n=1}^{\infty}nf_b(n)\left(\frac{\mu^2}{4\delta^2}+1\right)^{n-1}\eta^{n-1}\\
	&=\frac{\mathcal{G}_b(\eta;\mu,\delta)}{\sqrt{1-\eta}},
\end{eqnarray}
where
\begin{equation}
	\mathcal{G}_b(\eta;\mu,\delta)=\left[\frac{2\delta^2}{\sqrt{\mu^2+4\delta^2}}+b\left(1-\sqrt{1-\eta}\right)\right]\frac{\rme^{-\left[\sqrt{(\mu^2+4\delta^2)(1-\eta)}-\mu\right]\case{b}{2\delta^2}}}{\sqrt{\mu^2+4\delta^2}-\mu}.
\end{equation}
In this way, for $b$ fixed the rhs has a singular behaviour as $\eta\to1$ of the form $\mathcal{G}_b(1;\mu,\delta)/\sqrt{1-\eta}$ and consequently a use of Tauberian theorems yields
\begin{equation}
	f_b(n)\sim\frac{(2\delta^2+b\sqrt{\mu^2+4\delta^2})\rme^{\case{\mu b}{2\delta^2}}}{\sqrt{\mu^2+4\delta^2}(\sqrt{\mu^2+4\delta^2}-\mu)}\frac{(\frac{\mu^2}{4\delta^2}+1)^{1-n}}{\sqrt{4\pi n^3}}.
\end{equation}
Hence, for both $\mu>0$ and $\mu<0$, $f_b(n)$ displays an exponential decay modulated by a power-law. We note that the coefficient of the decay is larger when $\mu>0$. If we divide $f_b(n)$ by $\mathcal{E}_b$, we obtain
\begin{equation}
	\frac{f_b(n)}{\mathcal{E}_b}\sim\frac{(2\delta^2+b\sqrt{\mu^2+4\delta^2})\rme^{\case{|\mu| b}{2\delta^2}}}{\sqrt{\mu^2+4\delta^2}(\sqrt{\mu^2+4\delta^2}-|\mu|)}\frac{(\frac{\mu^2}{4\delta^2}+1)^{1-n}}{\sqrt{4\pi n^3}},
\end{equation}
in agreement with \eref{eq:Skw_Laplace_F_over_E}.

\paragraph{Leap-over.} When the walk enters $[b,\infty)$ for the first time, it occupies a random position, given by $S_{\FPT{b}}$. The distance of $S_{\FPT{b}}$ from $b$ is called the leap-over $L_b$ and its characteristic function is $\FT{\mathcal{L}}_b(k)=\rme^{-\rmi kb}\mathcal{F}_b(k,1)$. The Fourier inversion in our example can be performed easily, yielding the exact leap-over distribution:
\begin{eqnarray}
	\mathcal{L}_b(\ell)&=\int_{-\infty}^{\infty}\frac{dk}{2\pi\rmi}\rme^{-\rmi k(\ell+b)}\mathcal{F}_b(k,1)\\
	&=\frac{\mathcal{E}_b}{2\delta^2}\left(\sqrt{\mu^2+4\delta^2}-\mu\right)\rme^{-\left(\sqrt{\mu^2+4\delta^2}-\mu\right)\case{\ell}{2\delta^2}}=\frac{\mathcal{E}_b}{\mu^+}\rme^{-\ell/\mu^+}.\label{eq:Skw_Laplace_leap}
\end{eqnarray}
In this case, the conditional distribution $\mathcal{L}_b(\ell)/\mathcal{E}_b$ has no dependence on $b$. However, this property is far from being general, as we will see later in \sref{s:General_skw} and \sref{s:Applications}.

\begin{figure}
	\centering
	\begin{subfigure}{0.485\textwidth}
		\centering
		\includegraphics[width=\textwidth]{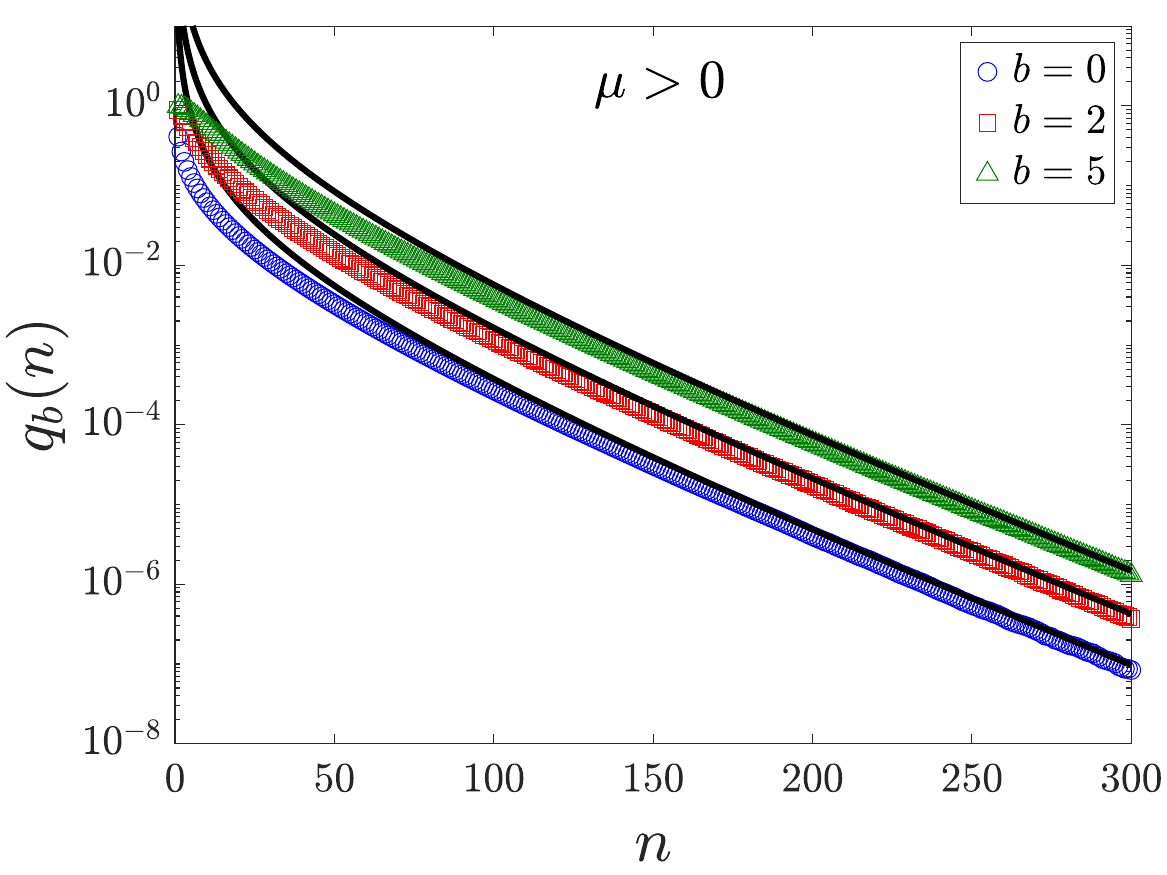}
		\caption{}
	\end{subfigure}
	\hfill
	\begin{subfigure}{0.485\textwidth}
		\centering
		\includegraphics[width=\textwidth]{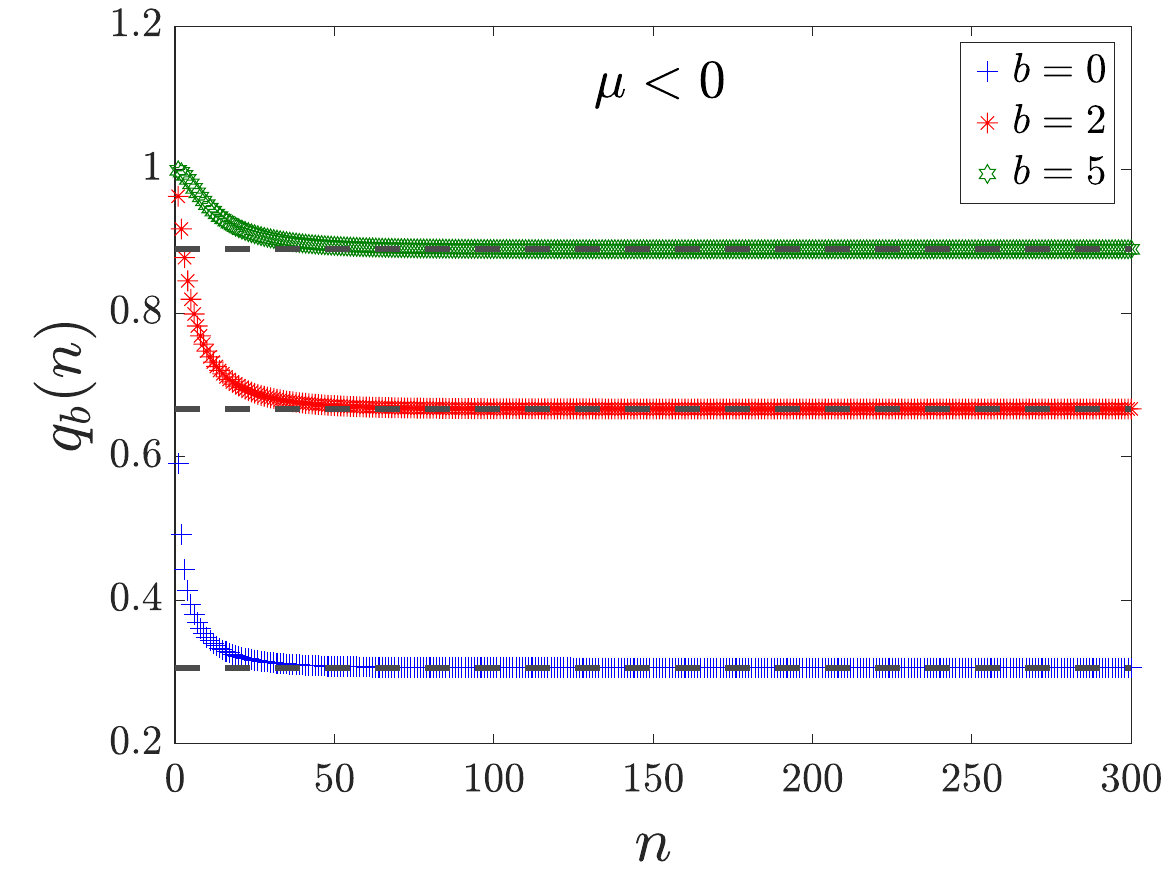}
		\caption{}
	\end{subfigure}
	\caption{Asymptotic decay of the survival probability for the Skewed Laplace distribution, for (a) $\mu>0$ and (b) $\mu<0$. The data are the results of numerical simulations, obtained by evolving (a) $10^9$ and (b) $10^8$ random walks of $300$ steps. Each jump is the sum of two exponential random variables, one positive with mean $\mu^+$ and one negative with mean $\mu^-=-1/\mu^+$. Thus, $\delta=\sqrt{-\mu^+\mu^-}=1$. In panel (a), we chose $\mu^+=1.2$, so that $\mu=\mu^++\mu^-=\case{11}{30}$. In panel (b) we chose instead $\mu^-=-1.2$, so that $\mu=-\case{11}{30}$. The asymptotic behaviours predicted by \eref{eq:Surv_SkwLapl_mu_pos} [solid lines in panel (a)] and \eref{eq:Surv_SkwLapl_mu_neg} [dashed lines in panel (b)] are in good agreement.}
	\label{fig:Surv_asym_SkwLapl}
\end{figure}

\paragraph{Survival probability.} From the relation $\mathcal{Q}_b(0,\zeta)=[1-\mathcal{F}_b(0,\zeta)]/(1-\zeta)$ and the previous results, we can obtain the asymptotic behaviour of the survival probability $q_b(n)$. If $\mu=0$, for $\zeta\approx1$ and $b$ fixed we have the expansion
\begin{equation}
	\mathcal{F}_b(0,\zeta)\sim1-\left(1+\frac b\delta\right)\sqrt{1-\zeta}+o\left(\sqrt{1-\zeta}\right),
\end{equation}
thus we conclude immediately that
\begin{equation}
	q_b(n)\sim\frac{1+b/\delta}{\sqrt{\pi n}}.
\end{equation}
If $\mu>0$, the transform $\mathcal{F}_b(0,\zeta)$ is analytic at $\zeta=1$, so we use the same idea as before and set $\zeta=(\case{\mu^2}{4\delta^2}+1)\eta$. Without repeating the calculations, we find that as $\eta\to1$ and $b$ fixed, to leading order
\begin{equation}
	\sum_{n=1}^{\infty}nq_n(n)(\kappa^2+1)^{n-1}\eta^{n-1}\sim\frac{4\delta^2}{\mu^2}\left.\frac{\partial\mathcal{F}_b(0,\zeta)}{\partial\zeta}\right|_{\zeta=\left(\case{\mu^2}{4\delta^2}+1\right)\eta}
\end{equation}
whence
\begin{equation}\label{eq:Surv_SkwLapl_mu_pos}
	q_b(n)\sim\frac{4\delta^2}{\mu^2}f_b(n)=\frac{4\delta^2(2\delta^2+b\sqrt{\mu^2+4\delta^2})\rme^{\case{\mu b}{2\delta^2}}}{\mu^2\sqrt{\mu^2+4\delta^2}(\sqrt{\mu^2+4\delta^2}-\mu)}\frac{(\frac{\mu^2}{4\delta^2}+1)^{1-n}}{\sqrt{4\pi n^3}}.
\end{equation}
Finally, if $\mu<0$ we have $\mathcal{F}_b(0,1)=\mathcal{E}_b<1$, thus
\begin{equation}
	\mathcal{Q}_b(0,\zeta)\sim\frac{1-\mathcal{E}_b}{1-\zeta},
\end{equation}
meaning that $q_b(n)$ decays to a fixed value given by $1-\mathcal{E}_b$: 
\begin{equation}\label{eq:Surv_SkwLapl_mu_neg}
	\lim_{n\to\infty}q_b(n)=1-\mathcal{E}_b=1-\left(\frac{\sqrt{\mu^2+4\delta^2}+\mu}{\sqrt{\mu^2+4\delta^2}-\mu}\right)\rme^{\mu b/\delta^2}.
\end{equation}
Recalling that $q_b(n)$ corresponds to the distribution of the maximum, viz. $q_b(n)=\mathbb{P}(M_n\leq b)$,
we see that in this case $M_n$ reaches a stationary state in the large-$n$ limit: $\lim_{n\to\infty}\Pr(M_n\leq b)=1-\mathcal{E}_b$. In \fref{fig:Surv_asym_SkwLapl} we confirm these results with numerical simulations, finding indeed good agreement between theory and data.

\subsection{Explicit example: sum of an arbitrary number of exponential random variables}\label{s:General_skw}
The idea used to define the Skewed Laplace random variable can be easily generalized. For example, each $X_i$ may be defined as the sum of $m$ copies of a positive exponential random variables with mean $\mu^+>0$, and $m$ copies of a negative exponential random variables with mean $\mu^-<0$. Then,
\begin{equation}\label{eq:Skewed_Erlang_CF}
	\FT{\wp}(k)=\frac{1}{(1-\rmi\mu^+k)^m(1-\rmi\mu^-k)^m}=\frac{1}{(1-\rmi\mu k+\delta^2k^2)^m},
\end{equation}
where $\delta^2=-\mu^+\mu^-$ and $\mu=\mu^++\mu^-$. Note that the sum of $m$ copies of an exponential random variable with mean $\mu^+$ is an Erlang random variable of order $m$ and mean $m\mu^+$. Thus, \eref{eq:Skewed_Erlang_CF} generalizes the Skewed Laplace case, being the characteristic function of the sum of two Erlang random variables of order $m$, one positive and one negative, with mean $m\mu^+$ and $m\mu^-$, respectively. The factorization of
\begin{equation}
	1-\zeta\FT{\wp}(k)=\frac{(1-\rmi\mu k+\delta^2k^2)^m-\zeta}{(1-\rmi\mu k+\delta^2k^2)^m}
\end{equation}
can be performed explicitly. Indeed, recalling the Skewed Laplace case, the denominator can be factorized as
\begin{equation}
	(1-\rmi\mu k+\delta^2k^2)^m=\left[r_-(0)-\rmi\delta k\right]^m[r_+(0)+\rmi\delta k]^m,
\end{equation}
where $r_-(\zeta)$ and $r_+(\zeta)$ are precisely defined by \eref{eq:r_l} and \eref{eq:r_u}. To factorize the numerator, we compute its roots. By setting $z=1-\rmi\mu k+\delta^2k^2$, it is easy to see that the roots can be expressed as
\begin{equation}\label{eq:z_roots}
	z^{j}(\zeta)=\zeta^{1/m}\rme^{2\pi\rmi j/m},\quad j=0,1,\dots,m-1.
\end{equation}
To go back to the variable $k$ we apply the transformation
\begin{equation}\label{eq:z_to_k_transf}
	k=\frac{\rmi}{2\delta^2}\left[\mu\pm\sqrt{\mu^2+4\delta^2(1-z)}\right],
\end{equation}
which maps a single $z$ to two values of $k$. Thus, for each $z^j(\zeta)$ we obtain two corresponding roots, say $\kappa^j_+(\zeta)$ and $\kappa^j_-(\zeta)$, given by
\begin{equation}
	\kappa_\pm^j(\zeta)=\frac{\rmi}{2\delta^2}\left\{\mu\pm\sqrt{\mu^2+4\delta^2[1-z^j(\zeta)]}\right\}.
\end{equation}
Therefore, by introducing $ r^j_-(\zeta)\coloneq\rmi \delta\kappa_-^j(\zeta)$ and $r^j_+(\zeta)\coloneq-\rmi \delta\kappa_+^j(\zeta)$, the numerator can be factorized as
\begin{equation}
	(1-\rmi\mu k+\delta^2k^2)^m-\zeta=\prod_{j=0}^{m-1}\left[r^j_-(\zeta)-\rmi\delta k\right]\left[r^j_+(\zeta)+\rmi\delta k\right].
\end{equation}
One can show that $\kappa_+^j(\zeta)$ always lies in the upper half-plane and $\kappa_-^j(\zeta)$ always in the lower half-plane. Then, both $r^j_+(\zeta)$ and $r^j_-(\zeta)$ have positive real part and we can conclude that
\begin{equation}
	\Psi_0^+(z,\zeta)=\frac{\prod_{j=0}^{m-1}[r^j_-(\zeta)-\rmi\delta k]}{[r_-(0)-\rmi\delta k]^m},
\end{equation}
whence one can begin the study.

\begin{figure}
	\centering
	\begin{subfigure}{0.485\textwidth}
		\centering
		\includegraphics[width=\textwidth]{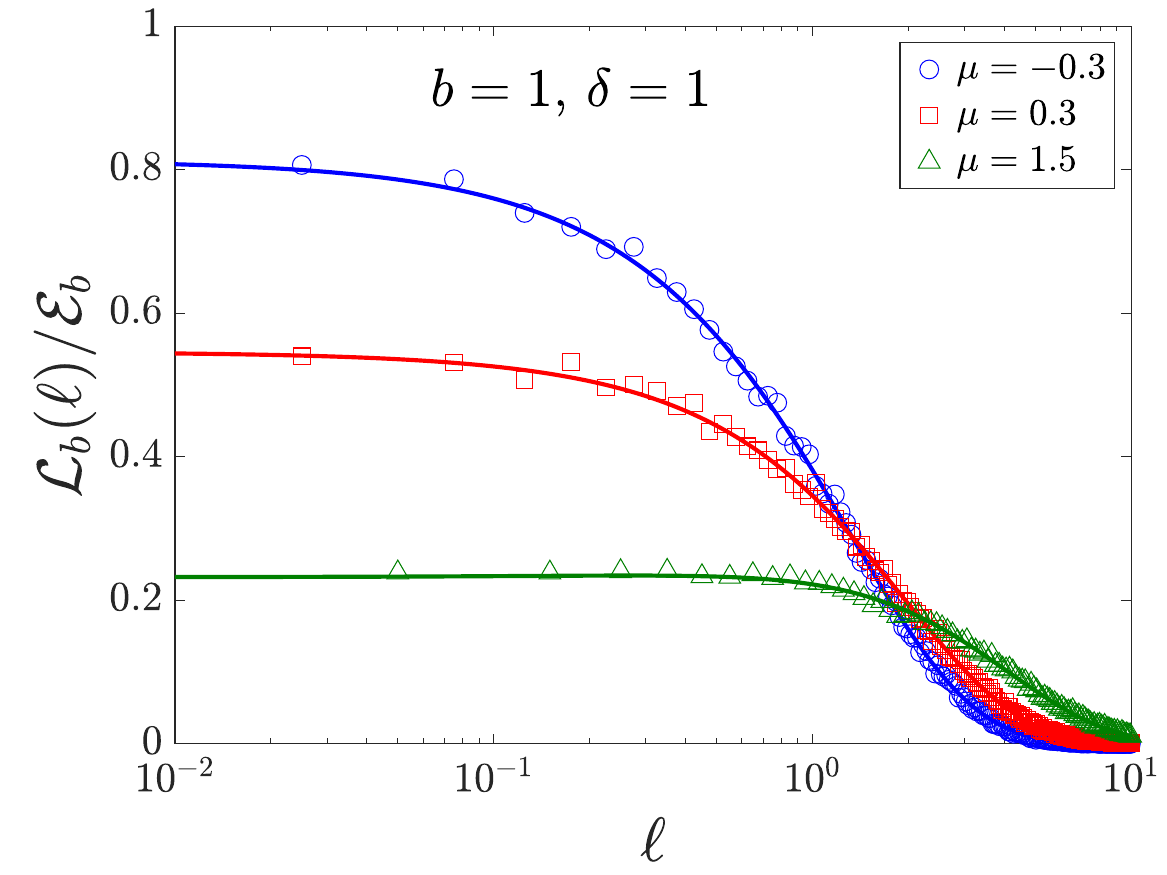}
		\caption{}
	\end{subfigure}
	\hfill
	\begin{subfigure}{0.485\textwidth}
		\centering
		\includegraphics[width=\textwidth]{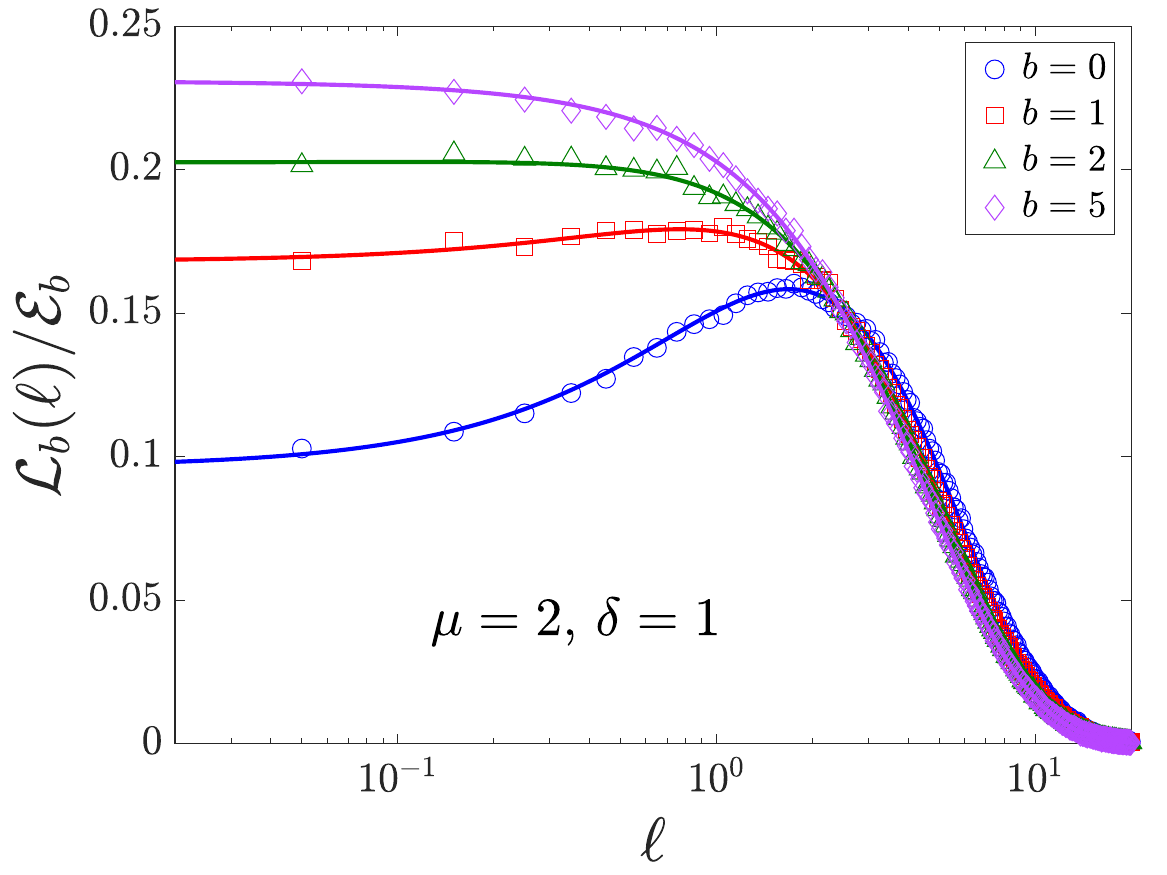}
		\caption{}
	\end{subfigure}
	\caption{Conditional leap-over PDF for the case $m=2$, given by the ratio \eref{eq:leap_biased_PDF} over \eref{eq:escape_biased}. Panel (a) shows distributions with different $\mu$ and $b$ fixed, while panel (b) displays different $b$ with $\mu$ fixed. The plots are in logarithmic scale on the $\ell$-axis to show that major differences are visible for small $\ell$ [see, in particular, panel (b)]. The data (markers) are the results of numerical simulations, obtained by evolving $N$ random walks up to the first-passage time in $(b,\infty)$ and measuring $L_b=S_{\FPT{b}}-b$ [panel (a): $N=10^5$; panel (b): $N=10^6$]. The numerical distributions are compared with the theoretical results of \eref{eq:leap_biased_PDF}, showing good agreement.}
	\label{fig:Leap_biased}
\end{figure}

As a rapid check, we recall that in the Skewed Laplace case, viz. $m=1$, there is a single root $z^0(\zeta)=\zeta$, and indeed one obtains $r^0_-(\zeta)=r_-(\zeta)$ as defined by \eref{eq:r_l}. Another case that is not too cumbersome to treat is $m=2$. There are two roots, $z^0(\zeta)=\sqrt{\zeta}$ and $z^1(\zeta)=-\sqrt{\zeta}$, and consequently,
\begin{eqnarray}
	r^0_-(\zeta)&=\frac{1}{2\delta}\left[\sqrt{\mu^2+4\delta^2(1-\sqrt{\zeta})}-\mu\right]\\
	r^1_-(\zeta)&=\frac{1}{2\delta}\left[\sqrt{\mu^2+4\delta^2(1+\sqrt{\zeta})}-\mu\right].
\end{eqnarray}
To simplify the following formulae, let $r=r_-(0)$, $r_1(\zeta)=r^0_-(\zeta)$ and $r_2(\zeta)=r^1_-(\zeta)$. Then,
\begin{equation}
	\Psi_0^+(z,\zeta)=\frac{[r_1(\zeta)-\rmi\delta k][r_2(\zeta)-\rmi\delta k]}{(r-\rmi\delta k)^2}.
\end{equation}
Again, the direct computation of $\Omega(k,\zeta)$ can be avoided, see \eref{eq:Skw_Laplace_Omega}. Finally, by employing \eref{eq:Hb}, we get
\begin{eqnarray}
	\mathcal{H}_b(k,\zeta)=\frac{1}{\Psi_0^+(k,\zeta)}&-\frac{[r-r_1(\zeta)]^2\rme^{\rmi kb-r_1(\zeta)b/\delta}}{[r_2(\zeta)-r_1(\zeta)][r_1(\zeta)-\rmi\delta k]}\nonumber\\
	&+\frac{[r-r_2(\zeta)]^2\rme^{\rmi kb-r_2(\zeta)b/\delta}}{[r_2(\zeta)-r_1(\zeta)][r_2(\zeta)-\rmi\delta k]}.
\end{eqnarray}
The resulting expression for $\mathcal{F}_b(k,\zeta)$ is
\begin{eqnarray}
	\mathcal{F}_b(k,\zeta)&=\frac{[r-r_1(\zeta)]^2[r_2(\zeta)-\rmi\delta k]}{[r_2(\zeta)-r_1(\zeta)](r-\rmi\delta k)^2}\rme^{\rmi kb-r_1(\zeta)b/\delta}\nonumber\\
	&\quad-\frac{[r_2(\zeta)-r]^2[r_1(\zeta)-\rmi\delta k]}{[r_2(\zeta)-r_1(\zeta)](r-\rmi\delta k)^2}\rme^{\rmi kb-r_2(\zeta)b/\delta},
\end{eqnarray}
and $\mathcal{Q}_b(k,\zeta)$ can then be obtained from \eref{eq:fund_eq_trans0}. We avoid the detailed analysis of the previous case and focus only on the leap-over. We recall that the leap-over characteristic function is given by $\FT{\mathcal{L}}_b(k)=\rme^{-\rmi kb}\mathcal{F}_b(k,1)$, which we can invert explicitly, obtaining:
\begin{eqnarray}
	\mathcal{L}_b(\ell)=&\Bigg\{\frac{(r-r_1(1))^2}{r_2(1)-r_1(1)}\left[\left(r_2(1)-r\right)\frac\ell\delta+1\right]\rme^{-r_1(1)b/\delta}\nonumber\\
	&+\frac{(r_2(1)-r)^2}{r_2(1)-r_1(1)}\left[\left(r-r_1(1)\right)\frac\ell\delta-1\right]\rme^{-r_2(1)b/\delta}\Bigg\}\frac{\rme^{-r\ell/\delta}}{\delta},\label{eq:leap_biased_PDF}
\end{eqnarray}
where
\begin{equation}
	r_2(1)=\frac{\sqrt{\mu^2+8\delta^2}-\mu}{2\delta},\quad r_1(1)=\frac{|\mu|-\mu}{2\delta},\quad r=\frac{\sqrt{\mu^2+4\delta^2}-\mu}{2\delta}.
\end{equation}
We remark that, since $\FT{\mathcal{L}}_b(0)=\mathcal{F}_b(0,1)=\mathcal{E}_b$, the total integral $\int_{-\infty}^{\infty}\mathcal{L}_b(\ell)d\ell=\mathcal{E}_b$, where
\begin{equation}\label{eq:escape_biased}
	\mathcal{E}_b=\frac{(r-r_1(1))^2r_2(1)}{(r_2(1)-r_1(1))r^2}\rme^{-r_1(1)b/\delta}-\frac{(r_2(1)-r)^2r_1(1)}{(r_2(1)-r_1(1))r^2}\rme^{-r_2(1)b/\delta}.
\end{equation}
It is straightforward to see that $\mathcal{E}_b=1$ for $\mu\geq0$, while for $\mu<0$ the escape probability depends on $\mu$. In \fref{fig:Leap_biased} we present the conditional distribution $\mathcal{L}_b(\ell)/\mathcal{E}_b$. Contrary to the case of the Skewed Laplace distribution, there is a clear dependence on $b$, see panel (b).

\subsection{Explicit example: Bernoulli random walk}
To complete the discussion, we treat a case with a discrete jump distribution. Let us consider a nearest-neighbour random walk with
\begin{equation}
	\wp(\xi)=p\delta(\xi+1)+q\delta(\xi-1),
\end{equation}
where $0<p<1$ and $q=1-p$. The function $\FT{\wp}_*(t)$, see \eref{eq:char_fun_*}, is
\begin{equation}
	\FT{\wp}_*(t)=\sum_{\xi\in\mathbb{Z}}t^\xi\wp(\xi)=pt+qt^{-1}.
\end{equation}
We first determine $\phi_0^+(w,\zeta)$ from \eref{eq:phi_0^+_in_main_formulae}. One can verify that $1-\zeta\FT{\wp}_*(t)$ can be factorized as
\begin{equation}
	1-\zeta\FT{\wp}_*(t)=\frac{p\zeta}{t}[t-w_I(\zeta)][w_E(\zeta)-t],
\end{equation}
where
\begin{eqnarray}
	w_I(\zeta)&=\frac{1}{2p\zeta}\left(1-\sqrt{1-4qp\zeta^2}\right)\\
	w_E(\zeta)&=\frac{1}{2p\zeta}\left(1+\sqrt{1-4qp\zeta^2}\right),
\end{eqnarray}
and $w_I(\zeta)$, $w_E(\zeta)$ are real numbers that lie in the interior and exterior of the unit circle, respectively. Consequently, $\ln[1-\zeta\FT{\wp}_*(t)]$ has three branch points, $w_0=0$, $w_I(\zeta)$ and $w_E(\zeta)$. In this case it is convenient to perform the integration directly, which yields
\begin{equation}
	\frac1{2\pi\rmi}\int_{\mathcal{C}}\frac{\ln[1-\zeta\FT{\wp}_*(t)]}{t-w}dt=\ln(p\zeta)+\ln[w_E(\zeta)-w],
\end{equation}
whence $\phi_0^+(w,\zeta)=p\zeta[w_E(\zeta)-w]$. This allows us to obtain the constant $B(\zeta)$ from \eref{eq:B}:
\begin{equation}
	B(\zeta)=\frac{1}{\sqrt{p\zeta w_E(\zeta)}}=\sqrt{\frac{1+\sqrt{4qp\zeta^2}}{4qp\zeta^2}}-\sqrt{\frac{1-\sqrt{4qp\zeta^2}}{4qp\zeta^2}}.
\end{equation}
We remark that for $q=p=\case12$ one recovers $B(\zeta)=(\sqrt{1+\zeta}-\sqrt{1-\zeta})/\zeta$, as found in \cite{MouMajSch-2020}.

As in the previous case, the direct computation of the phase $\omega(k,\zeta)$ can be avoided: as a consequence of the Sokhotski-Plemelj formulae [see \eref{eq:phi_0^pm} later in the paper], we have the relation
\begin{equation}
	\rme^{-\rmi\omega(k,\zeta)}=\frac{B(\zeta)\phi_0^+(\rme^{\rmi k},\zeta)}{\sqrt{1-\zeta\FT{\wp}(k)}}.
\end{equation}

Finally, to obtain $H_b(k,\zeta)$ we use \eref{eq:Hb_disc}. We note that $w=0$ is a pole of order $b+1$, hence:
\begin{eqnarray}
	H_b(k,\zeta)&=\phi_0^+(0,\zeta)\frac{\rme^{\rmi k(b+1)}}{p\zeta b!}\lim_{w\to0}\sum_{m=0}^{b}{b\choose m}\left[\frac1{w_E(\zeta)-w}\right]^{(m)}\left[\frac1{\rme^{\rmi k}-w}\right]^{(b-m)}\nonumber\\		&=\phi_0^+(0,\zeta)\frac{\rme^{\rmi k(b+1)}}{p\zeta b!}\lim_{w\to0}\sum_{m=0}^{b}{b\choose m}\frac{m!}{[w_E(\zeta)-w]^{m+1}}\frac{(b-m)!}{(\rme^{\rmi k}-w)^{b-m+1}}\nonumber\\	
	&=\frac{\phi_0^+(0,\zeta)}{p\zeta w_E(\zeta)}\frac{1-[\rme^{\rmi k}/w_E(\zeta)]^{b+1}}{1-\rme^{\rmi k}/w_E(\zeta)}=\frac{1-[\rme^{\rmi k}/w_E(\zeta)]^{b+1}}{1-\rme^{\rmi k}/w_E(\zeta)}.
\end{eqnarray}

By plugging these results in \eref{eq:Fb_disc} and \eref{eq:Qb_disc} and using $B(\zeta)=1/\sqrt{\phi_0^+(0,\zeta)}$, we obtain
\begin{eqnarray}
	\mathcal{F}_b(k,\zeta)&=\frac{\rme^{\rmi k(b+1)}}{w_E(\zeta)^{b+1}}\\
	\mathcal{Q}_b(k,\zeta)&=\frac{1}{[1-w_I(\zeta)\rme^{-\rmi k}]p\zeta w_E(\zeta)}\left[\frac{1-(\rme^{\rmi k}/w_E(\zeta))^{b+1}}{1-\rme^{\rmi k}/w_E(\zeta)}\right].
\end{eqnarray}

Again, we avoid a thorough analysing of these results. We just point out that the escape probability $\mathcal{E}_b$ is written in terms of $\mathcal{E}_0$ as
\begin{equation}
	\mathcal{E}_b=(\mathcal{E}_0)^{b+1},
\end{equation}
where
\begin{equation}
	\mathcal{E}_0=\mathcal{F}_0(0,1)=\frac1{w_E(1)}=\frac{2p}{1+|2p-1|}=\cases{
	1 & if $p\geq\case12$\\
	\frac{p}{1-p} & if $p<\case12$
	}
\end{equation}
If we introduce the mean jump $\langle X_i\rangle=p-q=\mu$, write $p=\case12+\case\mu2$ and $q=\case12-\case\mu2$ and compute $\mathcal{F}_b(0,\zeta)/\mathcal{E}_b$, we obtain
\begin{equation}
	\frac{\mathcal{F}_b(0,\zeta)}{\mathcal{E}_b}=\left[\frac{w_E(1)}{w_E(\zeta)}\right]^{b+1}=\zeta^{b+1}\left(\frac{1+|\mu|}{1+\sqrt{1-(1-\mu^2)\zeta^2}}\right)^{b+1},
\end{equation}
which is independent of the sign of $\mu$. Once again, this is a display of the first-passage duality.

\section{Asymptotic results: survival probability and leap-over for Lévy flights}\label{s:Applications}
In the previous section, we analysed in detail situations in which the peculiar form of the jump distribution leads to the derivation of closed-form results. Obviously, this is not possible in general. Nevertheless, the fact that our formulae (\ref{eq:Fb_cont})-(\ref{eq:Qb_cont}) and (\ref{eq:Fb_disc})-(\ref{eq:Qb_disc}) are written solely in term of the characteristic function of the jumps $\FT{\wp}(k)$ allows one to extract the asymptotic behaviour of the quantities of interest for a broad class of distributions, as we will show in this section.

\subsection{Survival probability generating function: breakdown of Sparre-Andersen universality}
The Sparre-Andersen theorem is an important result in fluctuation theory \cite{Sparre}. It states that for any random walk with iid jumps drawn from a continuous and symmetric distribution, the generating function $\mathcal{Q}_0(0,\zeta)$ is
\begin{equation}
	\mathcal{Q}_0(0,\zeta)=\frac{1}{\sqrt{1-\zeta}}
\end{equation}
independently of the fine details of $\FT{\wp}(k)$, as we can easily obtain from \eref{eq:Qb_cont}. It follows that, for a broad class of random walks, the survival probability $q_0(n)$ is exactly
\begin{equation}
	q_0(n)=\frac{\Gamma(n+\frac12)}{\Gamma(\frac12)\Gamma(n+1)},
\end{equation}
where $\Gamma(z)$ is the Gamma function. For large $n$, $q_0(n)$ displays an asymptotic power-law decay characterized by the exponent $\case12$, viz. $q_0(n)\sim 1/\sqrt{\pi n}$.

This remarkable universal form of the survival probability fails in the presence of asymmetric jump distributions. However, the deviations from the Sparre-Andersen result can be derived from \eref{eq:Qb_cont}.
Note that, in general, the characteristic function of an asymmetric distribution is a complex function, $\FT{\wp}(k)=\wpr(k)+\rmi\wpi(k)$. As a notable example, we analyse the case where the small-$k$ behaviour of $\wpr(k)$ is of the form $\wpr(k)\sim1-|ak|^\alpha$, where $a>0$ and $0<\alpha<2$, and the small-$k$ behaviour of $\wpi(k)$ is given by
\begin{equation}\label{eq:CF_Im_small-k}
	\wpi(k)\sim\cases{\beta\varpi(\alpha)\sgn(k)|ak|^\alpha+o(|k|^\alpha) & $0<\alpha<1$\\
		\mu k+o(k) & $\alpha=1$\\
		\mu k+\beta\varpi(\alpha)\sgn(k)|ak|^\alpha+o(|k|^\alpha)&$1<\alpha<2$}
\end{equation}
where $-1<\beta<1$, $-\infty<\mu<\infty$ and $\varpi(\alpha)=\tan(\pi\alpha/2)$. These expansions describe an asymmetric PDF $\wp(\xi)$ with diverging variance, as well as diverging mean in the range $0<\alpha\leq1$. Such kinds of random walks are called \emph{Lévy flights} and are widely studied in a broad range of applications. The effects of asymmetry on the survival probability of Lévy flights is studied in several works. For example, the role of a constant drift $\mu$ added to a symmetric jump distribution is investigated in \cite{LeDWie-2009,MajSchWer-2012,MouMajSch-2018,PadCapKan-2025}, while the impact of the distribution's skewness, measured by the parameter $\beta$, is examined in \cite{PadCheDyb-2019,PadCheDyb-2020}.

\begin{table}
	\caption{Asymptotic behaviours of the survival probability $q_0(n)$ for Lévy flights.}
	\label{tab:Table_surv}
	\begin{indented}
		\item[]\begin{tabular}{@{}||l|l|l|l||}
			\br
			$\alpha$&$\beta$&$\mu$&Large-$n$ behaviour of $q_0(n)$\\
			\br
			$(0,1)$&$(-1,1)$&any&$q_0(n)\propto n^{-\frac12-\frac1{\pi\alpha}\arctan[\beta\tan(\frac{\pi\alpha}{2})]}$\\
			\mr
			$1$&$0$&any&$q_0(n)\propto n^{-\frac12-\frac1{\pi}\arctan(\frac \mu a)}$\\
			\mr
			&&0&$q_0(n)\propto n^{-\frac12-\frac1{\pi\alpha}\arctan[\beta\tan(\frac{\pi\alpha}{2})]}$\\
			$(1,2)$&$(-1,1)$&positive&$q_0(n)\propto n^{-\alpha}$\\
			&&negative&$q_0(n)\to\mathrm{Const.}$\\
			\br
		\end{tabular}
	\end{indented}
\end{table}

\begin{figure}
	\centering
	\begin{subfigure}{0.485\textwidth}
		\centering
		\includegraphics[width=\textwidth]{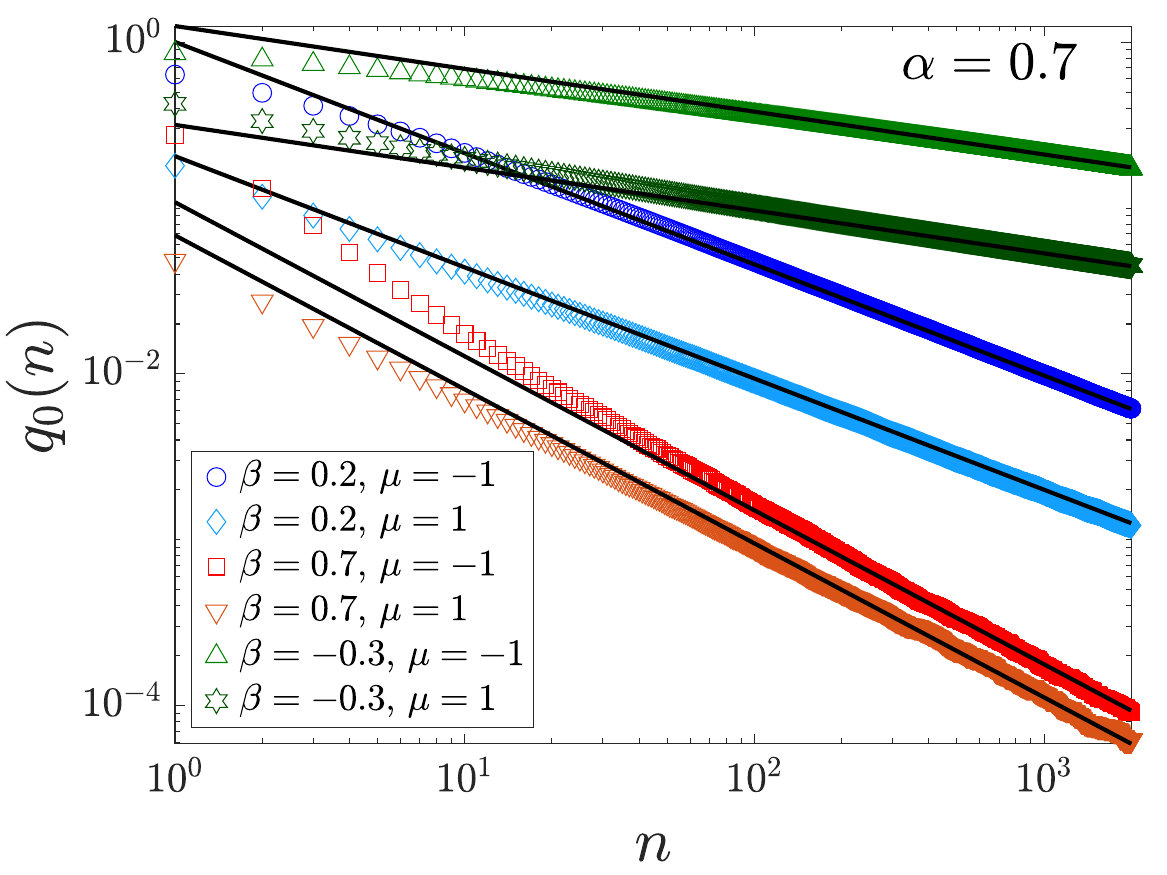}
		\caption{}
	\end{subfigure}
	\hfill
	\begin{subfigure}{0.485\textwidth}
		\centering
		\includegraphics[width=\textwidth]{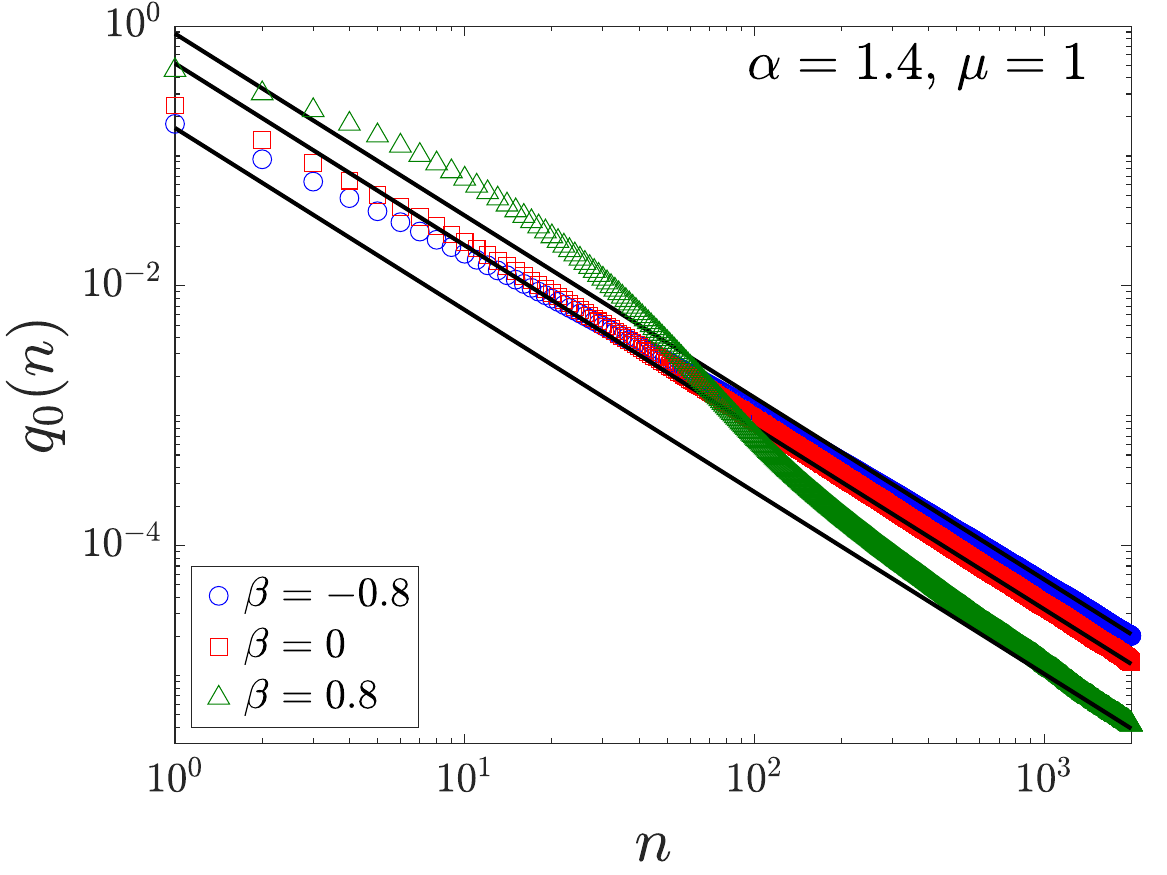}
		\caption{}
	\end{subfigure}
	\vfill
	\begin{subfigure}{0.485\textwidth}
		\centering
		\includegraphics[width=\textwidth]{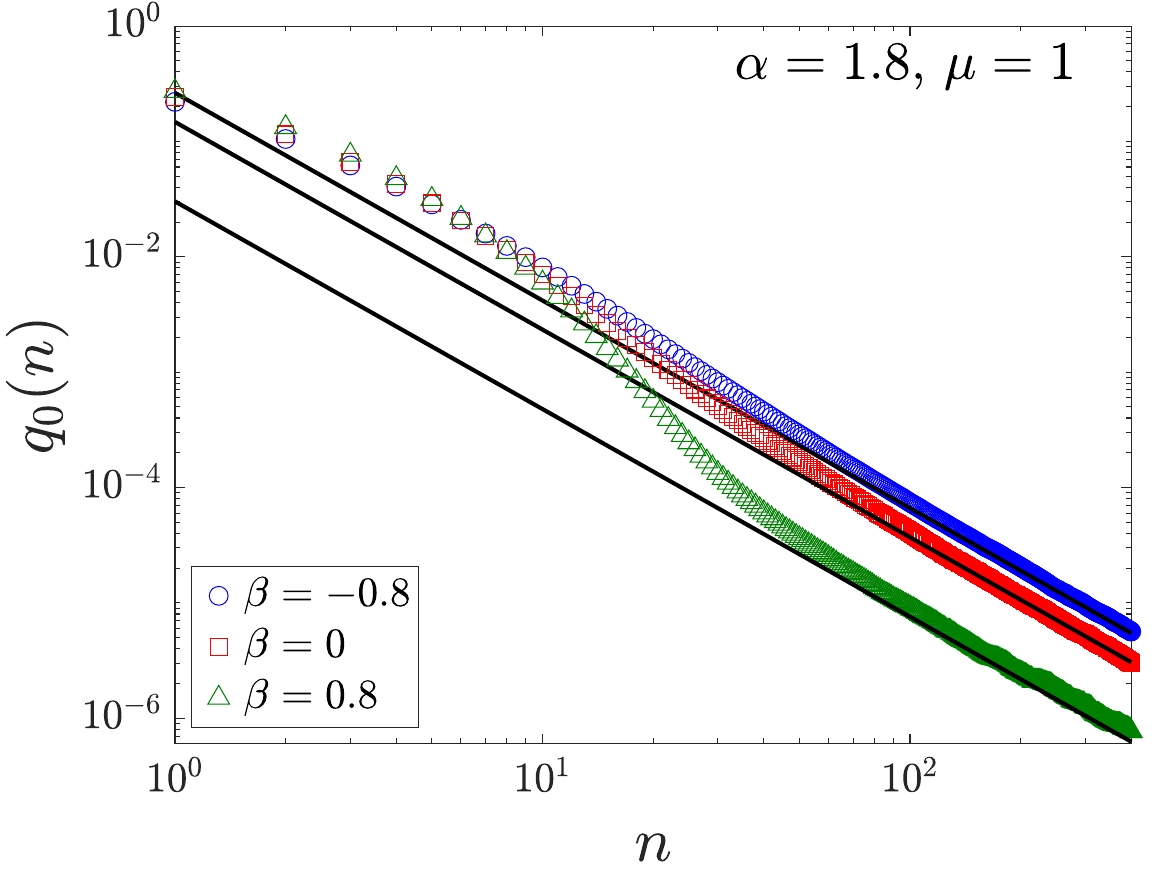}
		\caption{}
	\end{subfigure}
	\hfill
	\begin{subfigure}{0.485\textwidth}
		\centering
		\includegraphics[width=\textwidth]{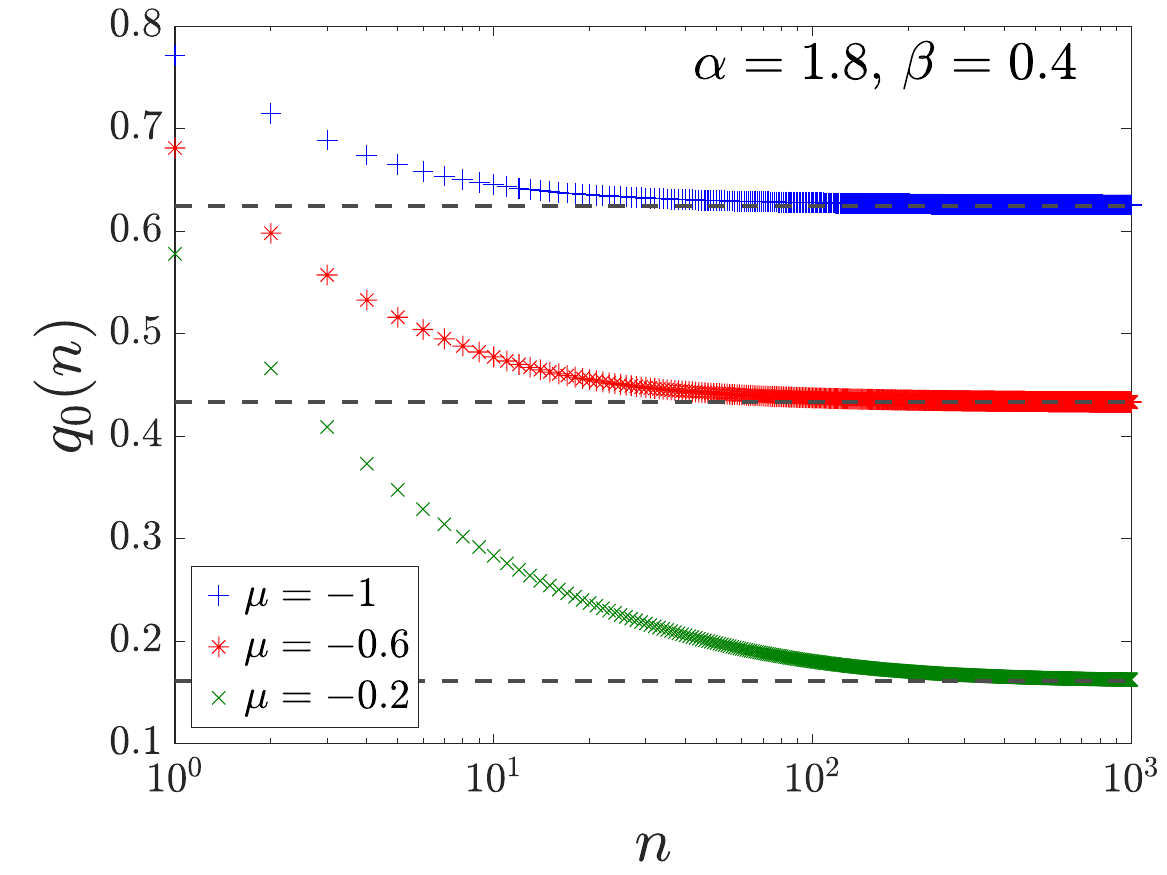}
		\caption{}
	\end{subfigure}
	\caption{Asymptotic decay of $q_0(n)$ for Lévy flights with jumps drawn from skewed Lévy stable laws. The asymptotic behaviour of the data is compared with our analytical predictions, see below. Panel (a): case $0<\alpha<1$. The data (markers) are obtained by evolving $10^6$ random walks and the analytical prediction (solid lines) are given by \eref{eq:App_q0_LS_a<1}. Panel (b)-(c): case $1<\alpha<2$, with $\mu>0$. The number of simulations is $10^8$. The data are compared with the analytical predictions of \eref{eq:App_q0_a>1_mu>0}. Panel (d): case $1<\mu<2$, with$\mu<0$. The number of simulation is $10^6$. The data converge to a constant, whose theoretical value (dashed line) is given by \eref{eq:App_q0_a>1_mu<0}.}
	\label{fig:Surv_asym}
\end{figure}

\begin{figure}
	\centering
	\includegraphics[width=0.6\textwidth]{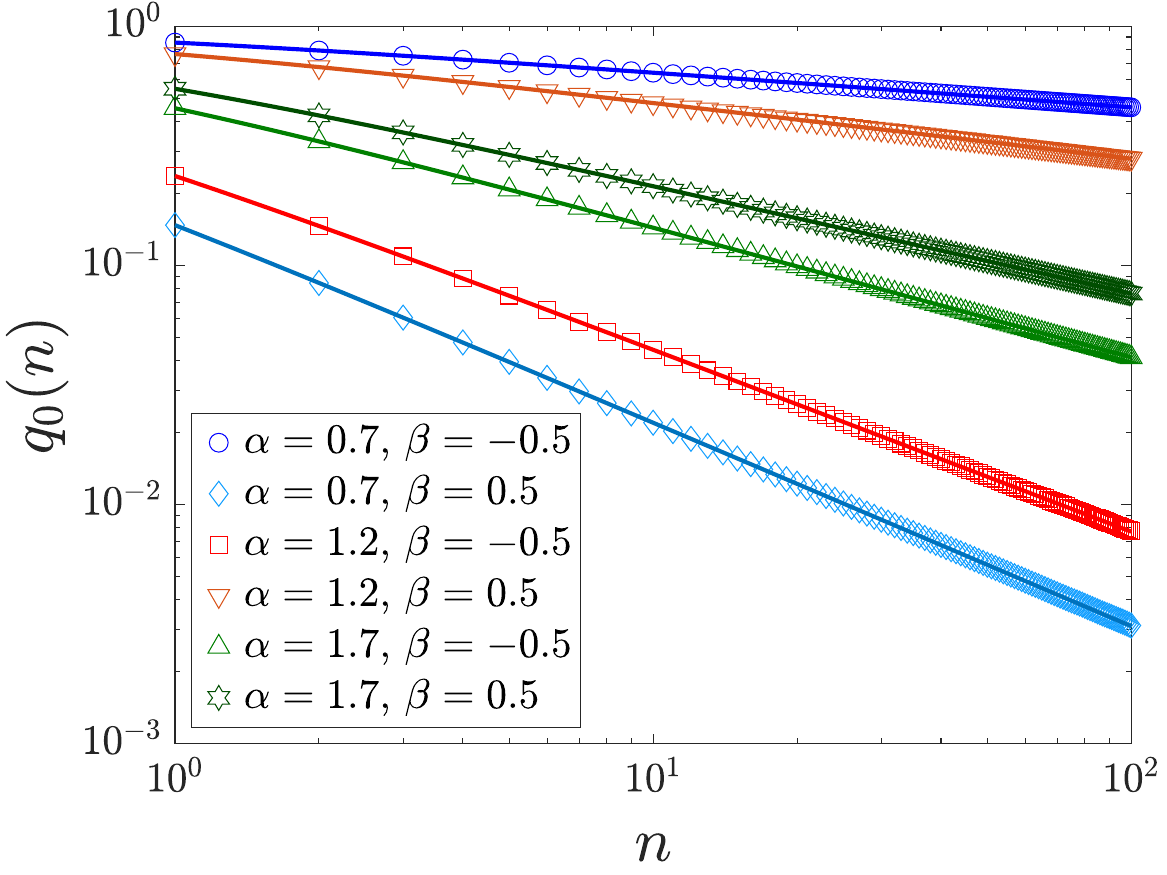}
	\caption{Exact survival probability $q_0(n)$ for Lévy flights, with jumps drawn from skewed Lévy stable laws with $\mu=0$. The exact theoretical result of \eref{eq:Surv_Levy_exact} is compared with our numerical simulations (markers). Each value of the survival probability is obtained by evolving $10^7$ random walks.}\label{fig:Surv_exact}
\end{figure}

In this work, we consider the combined effect of $\beta$ and $\mu$, and compute the exact asymptotic decay of $q_0(n)$ for $0<\alpha<2$. The decay exponents are consistent with those obtained in previous work \cite{MajSchWer-2012,PadCheDyb-2019} and are summarized in \tref{tab:Table_surv}. Furthermore, we obtain exact formulae for the coefficients, which are given only in terms of $\FT{\wp}(k)$. The technical details of the derivation are reported in \ref{s:App_LF_q}. We test these results by considering jumps drawn from from a Lévy stable law, with \cite{Sat}
\begin{equation}\label{eq:Lévy_CF}
	\FT{\wp}(k)=\cases{\exp\left\{\rmi \mu k-|ak|^\alpha\left[1-\rmi\beta\sgn (k)\varpi(\alpha)\right]\right\}&$\alpha\neq1$\\
	\exp\left(\rmi \mu k-|ak|\right)&$\alpha=1$}
\end{equation}  
which allows us to compute the coefficients explicitly. The results and the comparisons with numerical data are shown in \fref{fig:Surv_asym}. Interestingly, in the case $\mu=0$ we obtain that $q_0(n)$ is given exactly by
\begin{equation}\label{eq:Surv_Levy_exact}
	q_0(n)=\frac{\Gamma(n+\gamma)}{\Gamma(\gamma)\Gamma(n+1)},\quad\gamma=\frac12+\frac1{\pi\alpha}\arctan\left[\beta\tan\left(\frac{\pi\alpha}{2}\right)\right],
\end{equation}
see \fref{fig:Surv_exact}.

We remark that the case $\alpha=2$ corresponds to distributions possessing a finite variance $\sigma^2$. It was shown in \cite{MajSchWer-2012} that for such distributions the survival probability displays an exponential decay modulated by a power-law $\propto n^{-3/2}$ for $\mu>0$, while it decays to a constant for $\mu<0$. This is consistent with the explicit example we treated in \sref{s:Examples}, see equations \eref{eq:Surv_SkwLapl_mu_pos} and \eref{eq:Surv_SkwLapl_mu_neg}.

\subsection{Leap-over of symmetric Lévy flights.}
\begin{figure}
	\centering
	\begin{subfigure}{0.485\textwidth}
		\centering
		\includegraphics[width=\textwidth]{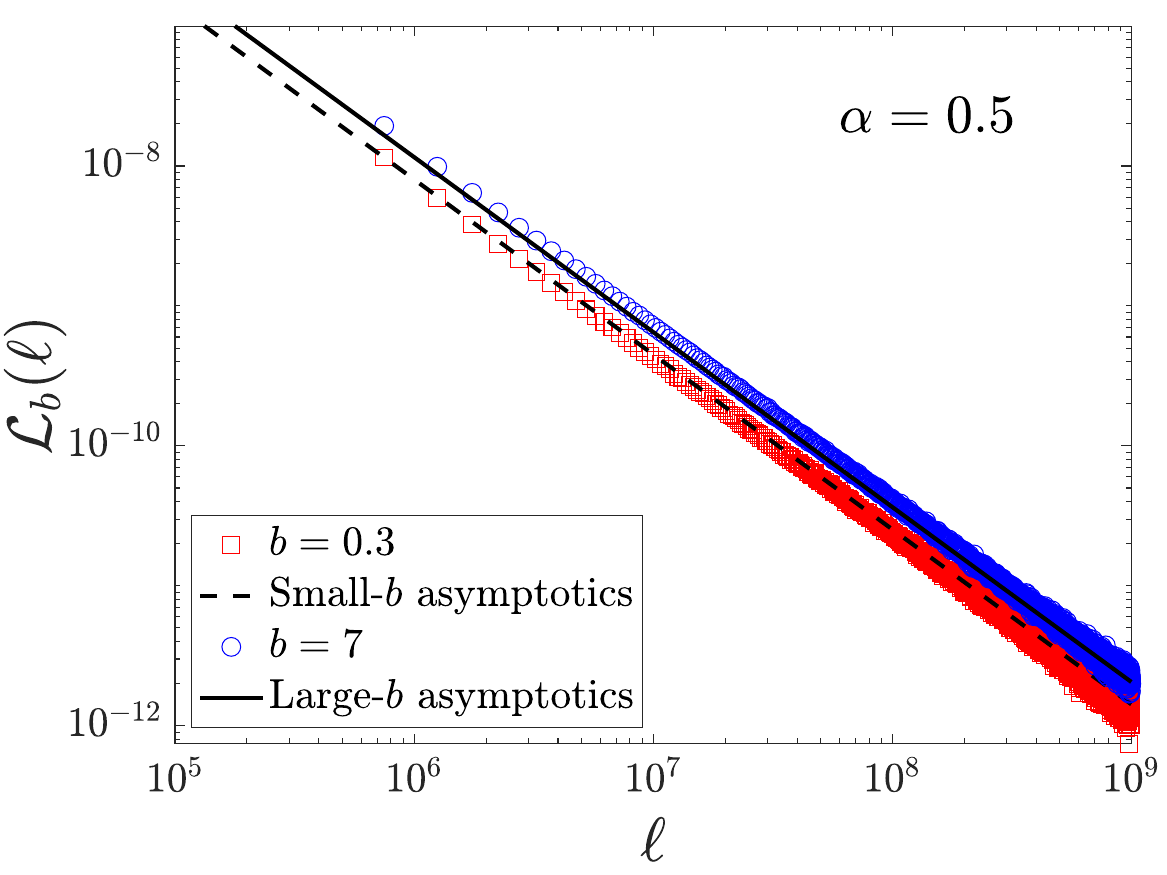}
		\caption{}
	\end{subfigure}
	\hfill
	\begin{subfigure}{0.485\textwidth}
		\centering
		\includegraphics[width=\textwidth]{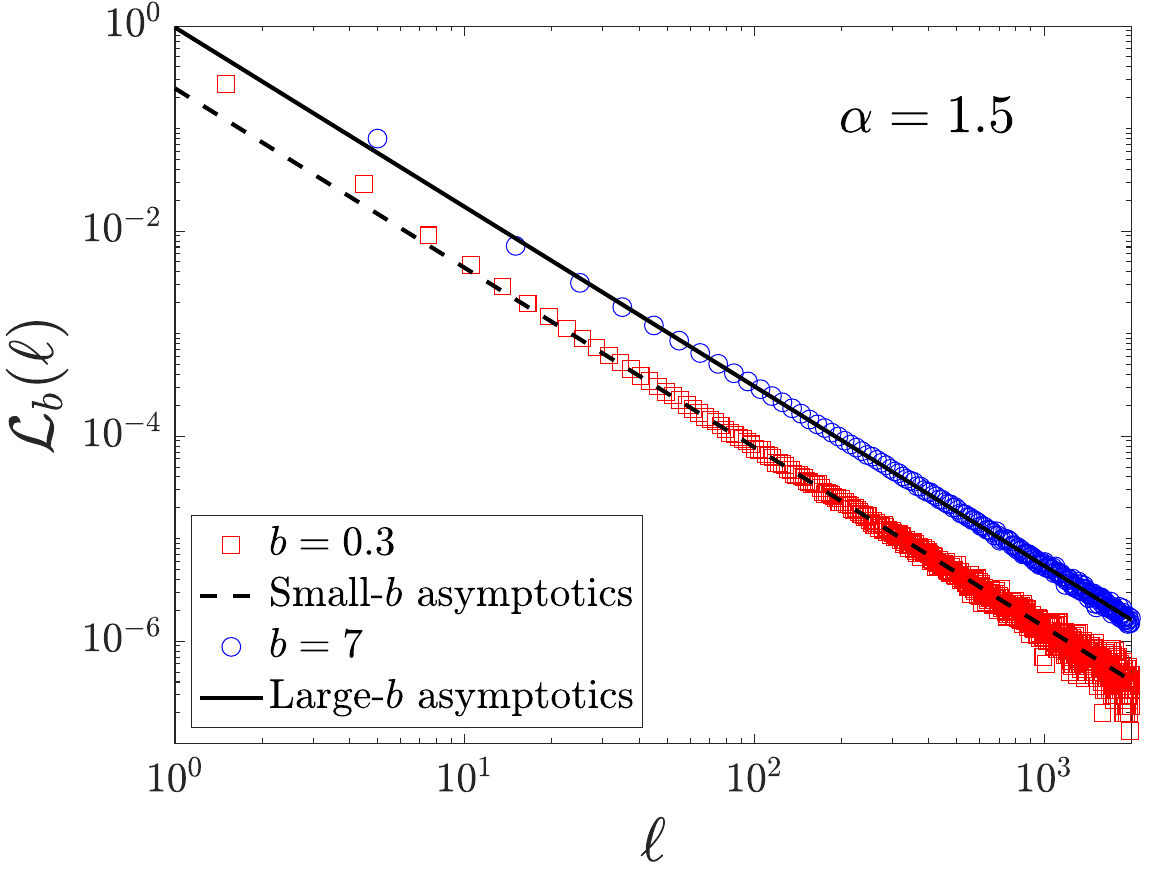}
		\caption{}
	\end{subfigure}
	\caption{Asymptotic power-law decay of the leap-over PDF for symmetric Lévy flights, with $\alpha=0.5$ [panel (a)] and $\alpha=1.5$ [panel (b)]. In both cases, the scale is $a=1$. The distributions are obtained by evolving $10^8$ [panel (a)] and $10^7$ [panel (b)] Lévy flights. The tails of the distributions are compared with the theoretical estimates given in \eref{eq:Leap_asympt}, displaying good agreement.}
	\label{fig:Leap_LF}
\end{figure}
The study of the leap-over has attracted particular interest in the literature when associated with Lévy flights \cite{Kor-Lom-al,PalCheMet-2014,Sin-1957,Ray-1958,Rog-1964}. Let us assume that $\wp(\xi)$ is symmetric, so that $\FT{\wp}(k)$ is real. Moreover, for small-$k$, $\FT{\wp}(k)\sim1-|ak|^\alpha$, with $0<\alpha<2$. By using the relation $\FT{\mathcal{L}}_b(k)=\rme^{-\rmi kb}\mathcal{F}_b(k,1)$ and evaluating the small-$k$ behaviour, we get
\begin{equation}\label{eq:Leap_PDF_Lévy_FT}
	\Re[\FT{\mathcal{L}}_b(k)]\sim1-\mathcal{H}_b(0,1)\cos(\pi\alpha/4)|ak|^{\alpha/2},
\end{equation}
see \ref{s:App_Leap_LF} for more details. Such a small-$k$ behaviour of the real part of the Fourier transform corresponds to the large-$\ell$ decay of the leap-over PDF:
\begin{equation}
	\mathcal{L}_b(\ell)\sim\frac{\alpha\mathcal{H}_b(0,1) a^{\alpha/2}}{2\Gamma(1-\alpha/2)}\ell^{-1-\alpha/2}.
\end{equation}
Remarkably, $\mathcal{L}_b(\ell)$ displays a slow power-law decay $\propto\ell^{-1-\alpha/2}$. Hence, for Lévy flights the distribution of $L_b$ is broader than the parent jump distribution and always possesses a diverging mean, even in the range $1<\alpha<2$ where the average jump length $\langle|X_1|\rangle$ is finite. The prefactor corresponds to the decay coefficient, whose dependence on $b$ is encompassed in the factor $\mathcal{H}_b(0,1)$.

We emphasize that, while in the continuous-time formalism the dependence of the distribution on $b$ is known exactly \cite{Kor-Lom-al,Ray-1958}, much less is known in the discrete-time scenario we consider here. Although in general one can not obtain $\mathcal{H}_b(0,1)$ explicitly, it is possible to obtain approximate expressions in the limits where the ratio $b/a$ is small or large, see \ref{s:App_Leap_LF}. We consequently get the following asymptotic behaviours:
\begin{equation}\label{eq:Leap_asympt}
	\mathcal{L}_b(\ell)\sim\frac{a^{\alpha/2}}{\ell^{1+\alpha/2}}\times\cases{
	\frac{\alpha I_0(2\sqrt{\lambda b/a})}{2\Gamma(1-\alpha/2)}&$\frac{b}{a}\to0$\\
	\frac{\sin(\pi\alpha/2)}{\pi}\left(\frac ba\right)^{\alpha/2}&$\frac{b}{a}\to\infty$
	}
\end{equation}
where $I_0(z)$ is the zero-order modified Bessel function of the first kind and $\lambda$ is a constant that depends on $\FT{\wp}(k)$, defined by
\begin{equation}\label{eq:Leap_lambda}
	\lambda\coloneq-\frac{a}{\pi }\int_{0}^{\infty}\ln[1-\FT{\wp}(k)]dk.
\end{equation}
For symmetric Lévy stable laws, with $\FT{\wp}(k)=\rme^{-|ak|^\alpha}$, the integral in \eref{eq:Leap_lambda} can be computed explicitly \cite{RADCri-2024}, yielding
\begin{equation}
	\lambda=\frac1{\pi}\zeta\left(1+\frac1\alpha\right)\Gamma\left(1+\frac1\alpha\right),
\end{equation}
where $\zeta(z)$ denotes the Riemann Zeta function. We point out that the case $\case ba\to\infty$ in \eref{eq:Leap_asympt} is precisely the dependence on $b$ obtained in the continuous-time formalism \cite{Kor-Lom-al,Ray-1958}. The different asymptotic regimes of \eref{eq:Leap_asympt} are illustrated in \fref{fig:Leap_LF} for two different values of $\alpha$ and compared with numerical simulations, showing good agreement with the theoretical predictions.

\section{Derivation of the main formulae: interpretation as a Riemann-Hilbert problem and solution}\label{s:derivation}
Here we derive the main formulae presented in \sref{s:Model}. As already mentioned, our approach starts from relation \eref{eq:fund_eq_trans0}. Let us introduce the complex variable $z=k+\rmi s$ and note that, according to the definitions \eref{eq:Fb_def} and \eref{eq:Qb_def}, the functions $\rme^{-\rmi zb}\mathcal{F}_b(z,\zeta)$ and $\rme^{-\rmi zb}\mathcal{Q}_b(z,\zeta)$ are bounded and analytic for $s>0$ and $s<0$, respectively. To take advantage of this, we multiply \eref{eq:fund_eq_trans0} by $\rme^{-\rmi kb}$ and then add $1$ on both sides, obtaining
\begin{equation}\label{eq:fund_eq_trans}
	1-\rme^{-\rmi kb}\mathcal{F}_b(k,\zeta)=\left[1-\zeta\FT{\wp}(k)\right]\rme^{-\rmi kb}\mathcal{Q}_b(k,\zeta)+1-\rme^{-\rmi kb}.
\end{equation}
The lhs of this equation can now be analytically extended to a bounded function in the upper half-plane, say $\Psi_b^+(z,\zeta)$. Similarly, $\rme^{-\rmi kb}\mathcal{Q}_b(k,\zeta)$ can be analytically extended to a bounded function in the lower half-plane, say $\Psi_b^-(z,\zeta)$. Together, they form the function $\Psi_b(z,\zeta)$ 
\begin{equation}\label{eq:Psi_b_cont}
	\Psi_b(z,\zeta)=\cases{
		\Psi_b^+(z,\zeta)\coloneq1-\rme^{-\rmi zb}\mathcal{F}_b(z,\zeta) &$\Im(z)>0$\\
		\Psi_b^-(z,\zeta)\coloneq\rme^{-\rmi zb}\mathcal{Q}_b(z,\zeta) &$\Im(z)<0$
	}
\end{equation}
which is bounded and analytic for $\Im(z)\neq0$. A function of this kind is called a \emph{sectionally analytic function}. In view of \eref{eq:fund_eq_trans}, the branches $\Psi_b^\pm(z,\zeta)$ satisfy for $\Im(z)=0$ the relation
\begin{equation}\label{eq:fund_eq_trans_Psi}
	\Psi_b^+(k,\zeta)=\left[1-\zeta\FT{\wp}(k)\right]\Psi_b^-(k,\zeta)+1-\rme^{-\rmi kb}.
\end{equation}
Note that, according to the definitions \eref{eq:Fb_def} and \eref{eq:Qb_def}, $\rme^{-\rmi zb}\mathcal{F}_b(z,\zeta)$ vanishes when $\Im(z)\to\infty$, while $\rme^{-\rmi zb}\mathcal{Q}_b(z,\zeta)$ converges to a constant for $\Im(z)\to-\infty$:
\begin{equation}\label{eq:C_b}
	\rme^{-\rmi zb}\mathcal{Q}_b(z,\zeta)\to C_b(\zeta)\coloneq\sum_{n=0}^{\infty}\zeta^n\mathbb{P}(S_n=b,\FPT{b}>n).
\end{equation}
This means that, since $\Psi_b(z,\zeta)$ is defined by \eref{eq:Psi_b_cont}, it must satisfy for $\Im(z)\to\pm\infty$
\begin{equation}\label{eq:Psi_b_infty}
	\Psi_b(z,\zeta)\to\cases{
		1 & for $\Im(z)\to+\infty$\\
		C_b(\zeta) & for $\Im(z)\to-\infty$
	}
\end{equation} 
Our aim is then to find a sectionally analytic function $\Psi_b(z,\zeta)$ that satisfies \eref{eq:Psi_b_infty} and whose branches $\Psi_b^\pm(z,\zeta)$ satisfy on the real line the linear relation \eref{eq:fund_eq_trans}.

The problem of reconstructing a function in the complex plane from a prescribed jump condition across a curve is known as \emph{Riemann-Hilbert problem} \cite{Gak,Abl-Fok,Its-2003}. The theory for this kind of problems is well known, so we will avoid a detailed discussion, which can be found e.g. in \cite{Gak}, and present only a brief general outline of method of solution. For the sake of clarity, let us first report more precisely the statement of a generic Riemann-Hilbert problem:

\emph{Suppose that $\mathcal{C}$ is a simple smooth closed contour, separating the complex plane into the interior domain $D^+$ (the domain within $\mathcal{C}$) and the exterior domain $D^-$ (the domain complementary to $D^++\mathcal{C}$). Two functions, $G(k)$ and $g(k)$, are given on $\mathcal{C}$, such that
	\begin{enumerate}
		\item $G(k)$ does not vanish on $\mathcal{C}$;
		\item \textbf{Hölder condition:} for any pair of points on $\mathcal{C}$, both $G(k)$ and $g(k)$ satisfy
		\begin{equation*}
			|f(k_1)-f(k_2)|\leq A|k_1-k_2|^\alpha,\quad A>0,\,\alpha>0;
		\end{equation*}
		\item \textbf{Decay condition:} if $\mathcal{C}$ contains $z=\infty$, then $G(k)$ tends to a well-defined limit $G(\infty)$ as $k\to\infty$, and for large $k$
		\begin{equation*}
			|G(k)-G(\infty)|<\frac{M}{|k|^\mu}\quad M>0,\,\mu >0.
		\end{equation*}
	\end{enumerate}
	Find two functions $\Psi^+(z)$ and $\Psi^-(z)$, which are analytic in $D^+$ and $D^-$, respectively, and satisfy on $\mathcal{C}$ the \textbf{boundary relation}
	\begin{equation*}\label{eq:Jump_condition}
		\Psi^+(k)=G(k)\Psi^-(k)+g(k).
	\end{equation*}
	In the particular case where $g(k)=0$, the problem is said \textbf{homogeneous}, otherwise \textbf{non-homogeneous}. The function $G(k)$ is called the coefficient of the problem and $g(k)$ its free term}.

To solve the problem, we follow these steps: first, we determine a particular special solution of the homogeneous problem, called the \emph{canonical solution}. Defining precisely what is the canonical solution is beyond the scope of this paper, one can refer to \cite{Gak} for more details. We assume here that $\ln G(k)$ is a single-valued function, in which case the canonical solution is a solution $\psi_0(z)$ that has no zeros, is analytic for $z\notin\mathcal{C}$, satisfies the condition at infinity $\psi_0(\infty)=1$ and the homogeneous boundary condition
\begin{equation}
	\psi_0^+(k)=G(k)\psi_0^-(k).
\end{equation}
This solution is given by
\begin{equation}
	\psi_0(z)=\exp\left[\frac{1}{2\pi\rmi}\int_{\mathcal{C}}\frac{\ln G(t)}{t-z}dt\right],\quad z\notin\mathcal{C}.
\end{equation}
The canonical solution is then used to construct the solution of the non-homogeneous problem. Note that $\psi_0^+(k)/\psi_0^-(k)=G(k)$, thus we can rewrite the boundary relation as
\begin{equation}\label{eq:jump_cond_mod}
	\frac{\Psi^+(k)}{\psi_0^+(k)}=\frac{\Psi^-(k)}{\psi_0^-(k)}+\frac{g(k)}{\psi_0^+(k)}.
\end{equation}
Now define
\begin{equation}
	I(z)\coloneq\frac{1}{2\pi\rmi}\int_{\mathcal{C}}\frac{g(t)}{\psi_0^+(t)}\frac{dt}{t-z},\quad z\notin\mathcal{C},
\end{equation}
and call $I^\pm(z)$ the branches of $I(z)$ when $z$ belongs to $D^+$ and $D^-$, respectively. By the Sokhotski-Plemelj formulae, see \ref{s:App_formule}, we have that for $k\in\mathcal{C}$
\begin{equation}
	I^+(k)-I^-(k)=\frac{g(k)}{\psi_0^+(k)},
\end{equation}
and we use this to rewrite \eref{eq:jump_cond_mod} as
\begin{equation}
	\frac{\Psi^+(k)}{\psi_0^+(k)}-I^+(k)=\frac{\Psi^-(k)}{\psi_0^-(k)}-I^-(k).
\end{equation}
The lhs can be analytically extended in the interior domain, and the rhs in the exterior domain. Thus, together they define an entire bounded function, hence a constant $A$. It follows that the general solution of the non-homogeneous Riemann-Hilbert problem is
\begin{equation}
	\Psi(z)=\psi_0(z)\left[I(z)+A\right].
\end{equation}

According to the discussion of \sref{s:Model}, it seems natural to identify $\mathcal{C}$ with the real line. With this choice, however, we need the coefficient of the problem $G(k)$ to satisfy the decay condition (iii). Since $G(k)=1-\zeta\FT{\wp}(k)$, with $0<\zeta<1$, the large-$k$ behaviour depends on the properties of $\FT{\wp}(k)$. If the jumps are continuous, the Riemann-Lebesgue Lemma ensures that $\FT{\wp}(k)$ decays to zero; on the other hand, if the jumps are discrete, $\FT{\wp}(k)$ generally does not tend to a limit, which suggests to make a different choice for $\mathcal{C}$. For this reason, in the following we will treat the two cases separately.

\subsection{Continuous jumps}
In this case $\FT{\wp}(k)$ is a Fourier transform
\begin{equation}
	\FT{\wp}(k)=\int_{-\infty}^{+\infty}\rme^{\rmi k\xi}\wp(\xi)d\xi.
\end{equation}
As already mentioned, the Riemann-Lebesgue Lemma allows us to make the natural choice $\mathcal{C}=(-\infty,\infty)$, and the canonical solution is thus written as
\begin{equation}\label{eq:psi_0}
	\psi_0(z,\zeta)=\exp\left\{\frac{1}{2\pi\rmi}\int_{-\infty}^{+\infty}\frac{\ln [1-\zeta\FT{\wp}(t)]}{t-z}dt\right\},\quad \Im(z)\neq0.
\end{equation}
We note that $\psi_0(z,\zeta)$ solves our Riemann-Hilbert problem in the case $b=0$. Indeed, setting $b=0$ in \eref{eq:fund_eq_trans_Psi} yields the homogeneous boundary relation
\begin{equation}
	\Psi_0^+(k,\zeta)=[1-\zeta\FT{\wp}(k)]\Psi_0^-(k,\zeta).
\end{equation}
Furthermore, since the jumps are continuous, for $b=0$ the constant $C_b(\zeta)$ is equal to $1$, see \eref{eq:C_b}. Thus the solution we seek converges to $1$ as $\Im(z)\to\pm\infty$, see \eref{eq:Psi_b_infty}, whence we conclude that $\Psi_0(z,\zeta)=\psi_0(k,\zeta)$. By using the Sokhotski-Plemelj formulae on \eref{eq:psi_0}, we find that the limit values $\Psi_0^\pm(k,\zeta)$ are given by
\begin{equation}\label{eq:psi_0_BV}
	\Psi_0^\pm(k,\zeta)=[1-\zeta\FT{\wp}(k)]^{\pm\frac12}\rme^{-\rmi\Omega(k,\zeta)},
\end{equation}
where $\Omega(k,\zeta)$ is defined by \eref{eq:Omega}. Consequently, we obtain the formulae for the transforms $\mathcal{F}_0(k,\zeta)$ and $\mathcal{Q}_0(k,\zeta)$:
\numparts
\begin{eqnarray}
	\mathcal{F}_0(k,\zeta)&=1-\Psi^+_0(k,\zeta)=1-\sqrt{1-\zeta\FT{\wp}(k)}\rme^{-\rmi\Omega(k,\zeta)}\\
	\mathcal{Q}_b(z,\zeta)&=\Psi_b^-(k,\zeta)=\frac{\rme^{-\rmi\Omega(k,\zeta)}}{\sqrt{1-\zeta\FT{\wp}(k)}}.
\end{eqnarray}
\endnumparts

To find the solution for $b>0$, we introduce the function
\begin{equation}
	I_b(z,\zeta)=\frac{1}{2\pi\rmi}\int_{-\infty}^{+\infty}\frac{1-\rme^{-\rmi tb}}{\Psi_0^+(t,\zeta)}\frac{dt}{t-z},\quad\Im(z)\neq0,
\end{equation}
so that the general solution is of the form
\begin{equation}\label{eq:gensol_cont}
	\Psi_b(z,\zeta)=\Psi_0(z,\zeta)\left[I_b(z,\zeta)+A_b(\zeta)\right].
\end{equation}
To determine the constant $A_b(\zeta)$, we use the condition at infinity \eref{eq:Psi_b_infty}. Note that for $b>0$ we have $C_b(\zeta)=0$, thus we seek $\Psi_b(z,\zeta)\to1$ as $\Im(z)\to\infty$ and $\Psi_b(z,\zeta)\to0$ as $\Im(z)\to-\infty$. By noting that
\begin{equation}
	\frac{1}{2\pi\rmi}\int_{-\infty}^{+\infty}\frac{1}{\Psi_0^+(t,\zeta)}\frac{dt}{t-z}=\cases{
		\frac{1}{\Psi_0^+(z,\zeta)}-\frac12 &for $\Im(z)>0$\\
		-\frac12 &for $\Im(z)<0$
	}
\end{equation}
we can rewrite $I_b(z,\zeta)$ as
\begin{equation}\label{eq:Ib_rewritten}
	I_b(z,\zeta)=\cases{
		\frac{1}{\Psi_0^+(z,\zeta)}-\frac12 -\frac{1}{2\pi\rmi}\int_{-\infty}^{+\infty}\frac{\rme^{-\rmi tb}}{\Psi_0^+(t,\zeta)}\frac{dt}{t-z}& $\Im(z)>0$\\
		-\frac12 - \frac{1}{2\pi\rmi}\int_{-\infty}^{+\infty}\frac{\rme^{-\rmi tb}}{\Psi_0^+(t,\zeta)}\frac{dt}{t-z}& $\Im(z)<0$
	}
\end{equation}
and observe that when $|z|\to\infty$, the integral at the rhs vanishes. Thus, recalling that $\Psi_0(z,\zeta)$ is the canonical solution, we obtain $I_b(z,\zeta)\to\case12$ as $\Im(z)\to\infty$ and $I_b(z,\zeta)\to-\case12$ as $\Im(z)\to-\infty$, whence we conclude $A_b(\zeta)=\case12$. Therefore, plugging $A_b(\zeta)=\case12$ and \eref{eq:Ib_rewritten} into \eref{eq:gensol_cont}, we find that the branches $\Psi_b^\pm(z,\zeta)$ are
\numparts
\begin{eqnarray}
	\Psi_b^+(z,\zeta)&=1-\frac{\Psi_0^+(z,\zeta)}{2\pi\rmi}\int_{-\infty}^{+\infty}\frac{\rme^{-\rmi tb}}{\Psi_0^+(t,\zeta)}\frac{dt}{t-z}\\
	\Psi_b^-(z,\zeta)&=-\frac{\Psi_0^-(z,\zeta)}{2\pi\rmi}\int_{-\infty}^{+\infty}\frac{\rme^{-\rmi tb}}{\Psi_0^+(t,\zeta)}\frac{dt}{t-z}.
\end{eqnarray}
\endnumparts
We can now obtain $\mathcal{F}_b(z,\zeta)$ and $\mathcal{Q}_b(z,\zeta)$ as follows:
\numparts
\begin{eqnarray}
	\mathcal{F}_b(z,\zeta)&=\rme^{\rmi zb}\left[1-\Psi^+_b(z,\zeta)\right]=\frac{\Psi_0^+(z,\zeta)}{2\pi\rmi}\int_{-\infty}^{+\infty}\frac{\rme^{-\rmi (t-z)b}}{\Psi_0^+(t,\zeta)}\frac{dt}{t-z}\\
	\mathcal{Q}_b(z,\zeta)&=\rme^{\rmi zb}\Psi_b^-(z,\zeta)=-\frac{\Psi_0^-(z,\zeta)}{2\pi\rmi}\int_{-\infty}^{+\infty}\frac{\rme^{-\rmi (t-z)b}}{\Psi_0^+(t,\zeta)}\frac{dt}{t-z}.
\end{eqnarray}
\endnumparts

The last step is to compute the limits when the imaginary part of $z=k+\rmi s$ tends to $0$ from above and below. The Sokhotski-Plemelj formulae yield
	\begin{eqnarray}
		\left[\int_{-\infty}^{+\infty}\frac{\rme^{-\rmi (t-z)b}}{\Psi_0^+(t,\zeta)}\frac{dt}{t-z}\right]^\pm&=\pm \frac{\pi\rmi}{\Psi_0^+(k,\zeta)}+\int_{-\infty}^{+\infty}\frac{\rme^{-\rmi (t-k)b}}{\Psi_0^+(t,\zeta)}\frac{dt}{t-k}\\ &=\pm\frac{\pi\rmi}{\Psi_0^+(k,\zeta)}-\int_{-\rmi\infty}^{+\rmi\infty}\frac{\rme^{ub}}{\Psi_0^+(k+\rmi u,\zeta)}\frac{du}{u},
\end{eqnarray}
where in the second line we first shifted the integration variable, $u=t-k$, and then performed a rotation, $u=\rmi u$. Note that at the rhs we have a singular integral. To compute its principal value, we consider a path $\mathcal{B}$ along the imaginary axis that passes on the right of the singularity $u=0$ along a semicircle of radius $\epsilon$ centred at the origin. The principal value is then obtained by subtracting the integral on the semicircle to the integral on $\mathcal{B}$ and taking the limit $\epsilon\to0$. Since the integral on the semicircle, in the limit, gives the contribution $\pi\rmi/\Psi_0^+(k,\zeta)$, we have
\begin{equation}
	\int_{-\rmi\infty}^{+\rmi\infty}\frac{\rme^{ub}}{\Psi_0^+(k+\rmi u,\zeta)}\frac{du}{u}=\int_{\mathcal{B}}\frac{\rme^{ub}}{\Psi_0^+(k+\rmi u,\zeta)}\frac{du}{u}-\frac{\pi\rmi}{\Psi_0^+(k,\zeta)},
\end{equation}
whence we obtain
\numparts
\begin{eqnarray}
	\mathcal{F}_b(k,\zeta)&=1-\Psi_0^+(k,\zeta)\mathcal{H}_b(k,\zeta)\label{eq:F_b_cont_vs_H}\\
	\mathcal{Q}_b(k,\zeta)&=\Psi_0^-(k,\zeta)\mathcal{H}_b(k,\zeta),
\end{eqnarray}
\endnumparts
where $\mathcal{H}_b(k,\zeta)$ is the integral defined in \eref{eq:Hb}. Equations \eref{eq:Fb_cont}-\eref{eq:Qb_cont} follow by using \eref{eq:psi_0_BV}. Finally, we prove that $\mathcal{H}_0(k,\zeta)=1$, so that the solutions are continuous with respect to $b$. Indeed, let us rewrite
\begin{equation}
	\mathcal{H}_b(k,\zeta)=\frac{1}{2\pi\rmi}\int_{_\mathcal{B}}\frac{\rme^{ub}}{u}du-\frac{1}{2\pi\rmi}\int_{_\mathcal{B}}\frac{\rme^{ub}}{u}\left[1-\frac{1}{\Psi_0^+(k+\rmi u,\zeta)}\right]du.
\end{equation}
The first integral at the rhs is equal to $1$ for any $b\geq0$. In the second integral instead we can take the limit $b\to0$ and then perform the integration by closing the contour with a semicircle $z=R\rme^{\rmi\theta}$, with $-\case\pi2<\theta<\case\pi2$. The contribution on the semicircle vanishes for $R\to\infty$, hence, by the residue theorem, the integral on $\mathcal{B}$ is zero, which proves the claim.

\subsection{Discrete jumps}
We assume that in this case the characteristic function $\FT{\wp}(k)$ is of the form
\begin{eqnarray}
	\FT{\wp}(k)&=\int\rme^{\rmi k\xi}\left[\sum_{m\in\mathbb{Z}}\, \mathbb{P}(X_1=am)\delta(\xi-a m)\right]d\xi\\
	&=\sum_{m\in\mathbb{Z}}\rme^{\rmi ka m}\, \mathbb{P}(X_1=a m),
\end{eqnarray}
where, without loss of generality, we can set $a=1$. Therefore, $\FT{\wp}(k)$ is a Fourier series, hence a periodic function that does not vanish for $|k|\to\infty$. To circumvent this issue, we will consider the problem over an interval of $k$ corresponding to a period.

Let $t=\rme^{\rmi k}$ be a point on the unit circle, with $k\in(-\pi,\pi)$. Furthermore, let us denote a generic complex number $w$ as $w=\rme^{\rmi z}$, with $z=k+\rmi s$. Note that $w$ lies in the interior of the unit circle if $s>0$ and in the exterior if $s<0$. Let us introduce the auxiliary sectionally analytic function
\begin{equation}\label{eq:Phi_b_cases}
	\Phi_b(w,\zeta)=\cases{
		\Phi_b^+(w,\zeta)\coloneq1-w^{-b}\mathcal{F}_b(-\rmi\ln w,\zeta) & $|w|<1$\\
		\Phi_b^-(w,\zeta)\coloneq w^{-b}\mathcal{Q}_b(-\rmi\ln w,\zeta) & $|w|>1$
	}
\end{equation}
Note that $\Phi_b^\pm(w,\zeta)$ are bounded and analytic in the interior and exterior of the unit circle, respectively. By using $t=-\rmi\ln k$ in \eref{eq:fund_eq_trans_Psi}, we obtain the boundary relation on the unit circle:
\begin{equation}
	\Phi_b^+(t,\zeta)=\left[1-\zeta\FT{\wp}_*(t)\right]\Phi_b^-(t,\zeta)+1-t^{-b},
\end{equation}
where $\FT{\wp}_*(t)=\FT{\wp}(-\rmi\ln t)$. In practice, we have replaced our original problem to an equivalent one where $\mathcal{C}$ is the unit circle. Since $\mathcal{C}$ does not contain the point at infinity, we can now disregard the decay condition. The conditions at infinity for $\Phi_b(w,\zeta)$ can be obtained from \eref{eq:Phi_b_cases}. We recall that $\rme^{-\rmi zb}\mathcal{F}_b(z,\zeta)$ vanishes when $\Im(z)\to\infty$, and $\rme^{-\rmi zb}\mathcal{Q}_b(z,\zeta)\to C_b(\zeta)$ when $\Im(z)\to-\infty$, which can be translated to
\begin{equation}
	\Phi_b(0,\zeta)=1,\quad\Phi_b(\infty,\zeta)=C_b(\zeta).
\end{equation}

We call $\phi_0(w,\zeta)$ the canonical solution of this problem, which is given by
\begin{equation}\label{eq:phi_0}
	\phi_0(w,\zeta)=\exp\left\{\frac1{2\pi\rmi}\int_{\mathcal{C}}\frac{\ln[1-\zeta\FT{\wp}_*(t)]}{t-w}dt\right\},\quad|w|\neq1.
\end{equation}
Following the same steps as before, we introduce
\begin{equation}\label{eq:Ib_discr}
	I_b(w,\zeta)=\frac{1}{2\pi \rmi}\int_{\mathcal{C}}\frac{1-t^{-b}}{\phi_0^+(t,\zeta)}\frac{dt}{t-w},\quad|w|\neq1,
\end{equation}
and write the general solution as
\begin{equation}\label{eq:gensol_disc}
	\Phi_b(w,\zeta)=\phi_0(w,\zeta)\left[I_b(w,\zeta)+A_b(\zeta)\right].
\end{equation}
To determine the constant $A_b(\zeta)$, we use the fact that $\Phi_b(\infty,\zeta)=C_b(\zeta)$, and that $I_b(\infty,\zeta)=0$. This implies $A_b(\zeta)=C_b(\zeta)$. Furthermore, from the condition $\Phi_b(0,\zeta)=1$ it follows $A_b(\zeta)=1-I_b(0,\zeta)$. This means that we can set
\begin{eqnarray}
	I_b(w,\zeta)+A_b(\zeta)&=1+I_b(w,\zeta)-I_b(0,\zeta)\nonumber\\
	&=\frac{1}{2\pi \rmi}\int_{\mathcal{C}}\frac{1-wt^{-b-1}}{\phi_0^+(t,\zeta)}\frac{dt}{t-w}.
\end{eqnarray}
where we used
\begin{equation}
	1=\frac{1}{2\pi \rmi}\int_{\mathcal{C}}\frac{1}{\phi_0^+(t,\zeta)}\frac{dt}{t}.
\end{equation}
Therefore,
\begin{equation}\label{eq:Phi_b}
	\Phi_b(w,\zeta)=\frac{\phi_0(w,\zeta)}{2\pi\rmi}\int_{\mathcal{C}}\frac{1-wt^{-b-1}}{\phi_0^+(t,\zeta)}\frac{dt}{t-w}.
\end{equation}

Before moving on, we first state some preliminary results. When $w$ is a point on the unit circle, viz. $w=\rme^{\rmi k}$, the boundary values $\phi_0^\pm(\rme^{\rmi k},\zeta)$ of the canonical solution are
\begin{eqnarray}
	\phi_0^\pm(\rme^{\rmi k},\zeta)&=[1-\zeta\FT{\wp}(k)]^{\pm\frac12}\exp\left\{\frac1{2\pi\rmi}\int_{\mathcal{C}}\frac{\ln[1-\zeta\FT{\wp}_*(t)]}{t-\rme^{\rmi k}}dt\right\}\\
	&=[1-\zeta\FT{\wp}(k)]^{\pm\frac12}\exp\left\{\frac1{2\pi}\int_{-\pi}^{\pi}\frac{\ln[1-\zeta\FT{\wp}(\theta)]}{1-\rme^{\rmi (k-\theta)}}d\theta\right\},\label{eq:tmp_1}
\end{eqnarray}
where in the second line we performed the change of variable $t=\rme^{\rmi\theta}$. By using the identity
\begin{equation}
	\frac{1}{1-\rme^{\rmi\varphi}}=\frac{1+\rmi\cot(\varphi/2)}{2},
\end{equation}
we can rewrite
\begin{eqnarray}
	\int_{-\pi}^{\pi}\frac{\ln[1-\zeta\FT{\wp}(\theta)]}{1-\rme^{\rmi (k-\theta)}}d\theta=&\frac12\int_{-\pi}^{\pi}\ln[1-\zeta\FT{\wp}(\theta)]d\theta\nonumber\\
	&-\frac{\rmi}{2}\int_{-\pi}^{\pi}\ln[1-\zeta\FT{\wp}(\theta)]\cot\left(\frac{\theta-k}{2}\right)d\theta,\label{eq:tmp_2}
\end{eqnarray}
where the first integral at the rhs is not singular and the second term is $ -2\pi\rmi\omega(k,\zeta)$, see \eref{eq:omega}. Therefore, by combining \eref{eq:tmp_1} and \eref{eq:tmp_2}, we find the relation
\begin{eqnarray}
	\phi_0^\pm(\rme^{\rmi k},\zeta)&=[1-\zeta\FT{\wp}(k)]^{\pm\frac12}\exp\left\{\frac1{4\pi}\int_{-\pi}^{\pi}\ln[1-\zeta\FT{\wp}(\theta)]d\theta-\rmi\omega(k,\zeta)\right\}\\
	&=[1-\zeta\FT{\wp}(k)]^{\pm\frac12}\frac{\rme^{-\rmi\omega(k,\zeta)}}{B(\zeta)}.\label{eq:phi_0^pm}
\end{eqnarray}
Now we note that in the case $b=0$, we have $I_0(w,\zeta)=0$, see \eref{eq:Ib_discr}, and $A_0(\zeta)=C_0(\zeta)$. Hence, according to \eref{eq:gensol_disc}, the solution is $\Phi_0(w,\zeta)=C_0(\zeta)\phi_0(w,\zeta)$. Furthermore, from the condition $\Phi_0(0,\zeta)=1$ it follows that $C_0(\zeta)=1/\phi_0(0,\zeta)$, and by using formula \eref{eq:phi_0} we obtain
\begin{eqnarray}
	C_0(\zeta)&=\exp\left\{-\frac1{2\pi\rmi}\int_{\mathcal{C}}\frac{\ln[1-\zeta\FT{\wp}_*(t)]}{t}dt\right\}\\
	&=\exp\left\{-\frac1{2\pi}\int_{-\pi}^{\pi}\ln[1-\zeta\FT{\wp}(\theta)]d\theta\right\}=B(\zeta)^2.
\end{eqnarray}
Hence, multiplying \eref{eq:phi_0^pm} by $C_0(\zeta)$ yields
\begin{equation}
	\Phi_0^\pm(\rme^{\rmi k},\zeta)=[1-\zeta\FT{\wp}(k)]^{\pm\frac12}B(\zeta)\rme^{-\rmi\omega(k,\zeta)}.\label{eq:Phi_0^pm}
\end{equation}
To obtain the transforms $\mathcal{F}_b(k,\zeta)$ and $\mathcal{Q}_b(k,\zeta)$, observe that
\begin{equation}
	\frac{1}{2\pi\rmi}\int_{\mathcal{C}}\frac{1}{\phi_0^+(t,\zeta)}\frac{dt}{t-w}=\cases{
		\frac{1}{\phi_0^+(w,\zeta)} &for $|w|<1$\\
		0 &for $|w|>1$
	}
\end{equation}
whence
\numparts
\begin{eqnarray}
	\mathcal{F}_b(-\rmi\ln w,\zeta)&=w^b[1-\Phi_b^+(w,\zeta)]=\phi_0^+(w,\zeta)J_b(w,\zeta)\\
	\mathcal{Q}_b(-\rmi\ln w,\zeta)&=w^b\Phi_b^-(w,\zeta)=-\phi_0^-(w,\zeta)J_b(w,\zeta),
\end{eqnarray}
\endnumparts
see \eref{eq:Phi_b}, where
\begin{equation}
	J_b(w,\zeta)=\frac{w^{b+1}}{2\pi\rmi}\int_{\mathcal{C}}\frac{t^{-b-1}}{\phi_0^+(t,\zeta)}\frac{dt}{t-w},\quad|w|\neq1.
\end{equation}
The limits of $J_b(w,\zeta)$ when $w$ approaches a point $\rme^{\rmi k}$ on the unit circle are
\begin{eqnarray}
	J_b^\pm(\rme^{\rmi k},\zeta)&=\pm\frac{1}{2\phi_0^+(\rme^{\rmi k},\zeta)}+\frac{\rme^{\rmi k(b+1)}}{2\pi\rmi}\int_{\mathcal{C}}\frac{t^{-b-1}}{\phi_0^+(t,\zeta)}\frac{dt}{t-\rme^{\rmi k}},
\end{eqnarray}
where, to compute the integral at the rhs, we use a strategy similar to the previous case. We consider the integral on a path $\LT{\mathcal{C}}$, which is the unit circle deformed to avoid $\rme^{\rmi k}$. This point is avoided by passing along a semicircle of radius $\epsilon$ centred on $\rme^{\rmi k}$, in such a way that $\rme^{\rmi k}$ is left in the exterior. Again, in the limit $\epsilon\to0$ the difference between the integral on $\LT{\mathcal{C}}$ and the integral on the semicircle yields the principal value of the singular integral. Since $\LT{\mathcal{C}}$ is defined in such a way that the only singular point in the interior is the origin, the former integral may be evaluated with the residue theorem, while the latter tends to the limit $-\rme^{-\rmi k(b+1)}/[2\phi_0^+(\rme^{\rmi k},\zeta)]$. By putting all together, we arrive at
\numparts
\begin{eqnarray}
	\mathcal{F}_b(k,\zeta)&=1-\phi_0^+(\rme^{\rmi k},\zeta)\mathrm{Res}\left[\frac{z^{-b-1}}{\phi_0^+(z,\zeta)}\frac{\rme^{\rmi k(b+1)}}{\rme^{\rmi k}-z},z=0\right]\\
	\mathcal{Q}_b(k,\zeta)&=\phi_0^-(\rme^{\rmi k},\zeta)\mathrm{Res}\left[\frac{z^{-b-1}}{\phi_0^+(z,\zeta)}\frac{\rme^{\rmi k(b+1)}}{\rme^{\rmi k}-z},z=0\right],
\end{eqnarray}
\endnumparts
which correspond to \eref{eq:Fb_disc}-\eref{eq:Qb_disc}. Note that for $b=0$
\begin{equation}
	\mathrm{Res}\left[\frac{z^{-1}}{\phi_0^+(z,\zeta)}\frac{\rme^{\rmi k}}{\rme^{\rmi k}-z},z=0\right]=\frac{1}{\phi_0^+(0,\zeta)},
\end{equation}
thus $H_0(k,\zeta)=1$, see \eref{eq:Hb_disc}.

\section{Conclusions and perspective}\label{s:Concl}
In this work, we studied first-passage problems for one-dimensional random walks with iid jumps starting from the origin. By mapping the initial problem to an equivalent Riemann-Hilbert problem, we were able to determine exact and semi-explicit general formulae for the quantities of interest, such as the joint distribution of the first-passage time and first-passage position beyond a threshold $b\geq0$, as well as the distribution of the random walks that do not cross $b$ up to step $n$, from which one can determine the survival probability in $(-\infty,b]$. Such formulae were used to carry out a thorough analysis of examples in which it was possible to obtain closed-form expressions, and an exact asymptotic analysis of examples for which closed-form results could not be obtained.
	
We point out that our approach can be applied to a more general setting, where the random walk $S_n$ is associated with a cost process $C_n$ \cite{BiaCriPoz-2025}. For instance, we can consider the case where each jump incurs a cost (e.g. time or energy), in which case one might be interested in the total cost incurred up to the first-passage time. Such a generalized random walk model has found recently many applications in physics \cite{MorMajViv-2024,MajMorViv-2023,MajMorViv-2023PRE,FucGolSei-2016,MorOlsKri-2023,SunBlyEva-2023,PalPalPar-2024,SunBlyEva-2025}. Our approach allows for exact results in cases involving asymmetries in the random walk or cost process, providing a deeper understanding of the model.

\ack
The work of MR has been partially supported by a research fellowship within the project ``Stochastic processes in non-homogeneous media as models of anomalous dynamics'', co-financed by BIR 2024, DIFA - DISAT Uninsubria - DM Unimib convention ``Models of anomalous dynamics in disordered and random media'' and PRIN 2017S35EHN funds of the Italian Ministry of Education, University and Research (MIUR).  This research is part of GC’s activity within GNFM/INdAM and the authors’ activity within the UMI Group “DinAmicI” (www.dinamici.org).

\appendix
\section{Cauchy principal value and Sokhotski-Plemelj formulae}\label{s:App_formule}
The results presented in this paper make use of the notion of Cauchy principal value of a singular integral and the Sokhostski-Plemelj formulae. Here we provide definitions for both.

\paragraph{Principal value of a singular integral:}Consider a definite integral of a real function $f(x)$ and suppose that $f(x)$ diverges at a point $c$ in the integration range, $a\leq c\leq b$. If
\begin{equation}
	\lim_{\epsilon\to0^+}\left[\int_{a}^{c-\epsilon}f(x)dx+\int_{c+\epsilon}^{b}f(x)dx\right]
\end{equation}
exists, the limit is called the Cauchy principal value of the integral. To avoid ambiguities, in this Appendix we denote it with the symbol $\mathrm{PV}\int$. The definition is based on the idea of avoiding the singularity by cutting out a symmetric neighbourhood of $c$. The same idea can be applied to the improper integral $\int_{-\infty}^{+\infty}f(x)dx$, which can be understood in the sense of the Cauchy principal value as
\begin{equation}
	\mathrm{PV}\int_{-\infty}^{+\infty}f(x)dx=\lim_{R\to\infty}\int_{-R}^{+R}f(x)dx,
\end{equation}
provided that the limit at the rhs exists. When the integration is taken over a curve $\mathcal{C}$, we take a small circle of radius $\epsilon$ centred on $c$, and denote as $\mathcal{C}-\epsilon$ the part of the curve cut out by the circle. The Cauchy principal value of the curvilinear integral is
\begin{equation}
	\mathrm{PV}\int_{\mathcal{C}}f(t)dt=\lim_{\epsilon\to0^+}\int_{\mathcal{C}-\epsilon}f(t)dt.
\end{equation}

\paragraph{Sokhotski-Plemelj formulae:}Let $\mathcal{C}$ be a simple smooth closed curve in the complex plane and $G(t)$ a function of position on $\mathcal{C}$, which is assumed to satisfy the Hölder condition on the curve. The function
\begin{equation}
	\psi(z)=\frac{1}{2\pi\rmi}\int_{\mathcal{C}}\frac{G(t)}{t-z}dt,
\end{equation}
defined by the Cauchy integral, is analytic for any $z\notin\mathcal{C}$. Furthermore, $\psi(z)$ splits into two independent analytic functions $\psi^\pm(z)$, depending on whether $z$ belongs to the interior domain $D^+$ or the exterior domain $D^-$. One can show that the functions $\psi^+(z)$ and $\psi^-(z)$ have limiting values $\psi^+(k)$ and $\psi^-(k)$ as $z$ approaches any point $k\in\mathcal{C}$ along any arbitrary path, and these values are given by the Sokhotski-Plemelj formulae:
\begin{eqnarray}
	\psi^+(k)=\frac12G(k)+\frac{1}{2\pi\rmi}\mathrm{PV}\int_{\mathcal{C}}\frac{G(t)}{t-k}dt\label{eq:App_SokPle+}\\
	\psi^-(k)=-\frac12G(k)+\frac{1}{2\pi\rmi}\mathrm{PV}\int_{\mathcal{C}}\frac{G(t)}{t-k}dt.
\end{eqnarray}
The same formulae apply when $\mathcal{C}$ is the real line, in which case $D^+$ corresponds to the upper-half plane and $D^-$ to the lower-half plane.

\section{Asymptotic behaviour of $q_0(n)$ for Lévy flights}\label{s:App_LF_q}
Starting from the general result
\begin{equation}\label{eq:App_Q0}
	\mathcal{Q}_0(0,\zeta)=\sum_{n=0}^{\infty}\zeta^nq_0(n)=\frac{\rme^{-\rmi\Omega(0,\zeta)}}{\sqrt{1-\zeta}},
\end{equation}
see \eref{eq:Qb_cont}, by employing Tauberian theorems one can deduce the large-$n$ behaviour of $q_0(n)$ from the singular behaviour of $\mathcal{Q}_0(0,\zeta)$ as $\zeta\to1$. To do so, we need to compute the singular behaviour of the phase $\Omega(0,\zeta)$. We first note that
\begin{eqnarray}\label{eq:App_id_IntAtan}
	-\rmi\Omega(0,\zeta)&=\frac{1}{2\pi\rmi}\int_{-\infty}^{\infty}\frac{dt}{t}\ln[1-\zeta\FT{\wp}(t)]\nonumber\\
	&=\frac{1}{2\pi}\int_{-\infty}^{\infty}\frac{dt}{t}\arg[1-\zeta\FT{\wp}(t)]\nonumber\\
	&=-\frac{1}{\pi}\int_{0}^{\infty}\frac{dt}{t}\arctan\left[\frac{\zeta\wpi(t)}{1-\zeta\wpr(t)}\right].
\end{eqnarray}
In the second line, we used the fact that $\ln[|1-\zeta\FT{\wp}(k)|]$ is an even function of $k$, thus the integrand is odd and thus the integral, understood in the sense of the principal value, vanishes. In the third line, we wrote explicitly $\arg[1-\zeta\wp(k)]$ and noted that the integrand is an even function. The integral is not singular because the divergence at $t=0$ is integrable for $0<\zeta<1$. Furthermore, by writing $\FT{\wp}(k)=\rho(k)\rme^{\rmi\vartheta(k)}$ and using the identity
\begin{equation}\label{eq:App_id_sum}
	\arctan\left[\frac{\Im(z)}{1-\Re(z)}\right]=\sum_{n=1}^{\infty}\frac{|z|^n}{n}\sin\left[n\arg(z)\right],\quad|z|\leq1,\, z\neq1,
\end{equation}
see also \cite{MajSchWer-2012}, we can rewrite
\begin{equation}\label{eq:App_Omega_sum}
	-\rmi\Omega(0,\zeta)=-\frac1\pi\sum_{n=1}^{\infty}\frac{\zeta^n}{n}\int_{0}^{\infty}\frac{\rho(t)^n}{t}\sin\left[n\vartheta(t)\right]dt.
\end{equation}
Equation \eref{eq:App_Omega_sum} will be particularly useful for the following analysis. We begin by considering Lévy stable laws and then expand to general heavy-tailed jump distributions.

\subsection{Lévy stable laws}
The exact characteristic function is explicitly given by \eref{eq:Lévy_CF}. Thus,
\begin{equation}
	\rho(k)=\rme^{-|ak|^{\alpha}},\quad\vartheta(k)=\cases{
	\mu k+\beta\varpi(\alpha)\sgn(k)|ak|^\alpha & $\alpha\neq1$\\
	\mu k &$\alpha=1$,
	}
\end{equation}
where $\varpi(\alpha)=\tan(\pi\alpha/2)$. First, we note that for $\alpha\neq1$, if we set $\mu=0$ the integral at the rhs of \eref{eq:App_Omega_sum} is
\begin{equation}
	\int_{0}^{\infty}\frac{\rme^{- t^\alpha}}{t}\sin[\beta\varpi(\alpha)t^\alpha]dt=\frac1\alpha\arctan\left[\beta\varpi(\alpha)\right],\label{eq:App_Int_LS_mu0}
\end{equation}
where first rescaled the integration variable $t\to t/a$ and then used the Frullani integral. By plugging this in \eref{eq:App_Omega_sum} and then using \eref{eq:App_Q0}, we obtain
\begin{equation}
	\mathcal{Q}_0(0,\zeta)=\frac1{(1-\zeta)^{1-\gamma}},\quad\gamma=\frac12+\frac1{\pi\alpha}\arctan\left[\beta\varpi(\alpha)\right],
\end{equation}
whence it follows the exact expression of $q_0(n)$ given in \eref{eq:Surv_Levy_exact}. With the same calculations, we can show that for $\alpha=1$ and any $\mu$,
\begin{equation}
	\int_{0}^{\infty}\frac{\rme^{-t}}{t}\sin\left(\frac{\mu t}{a}\right)dt=\arctan\left(\frac\mu a\right),
\end{equation}
thus
\begin{equation}
	\mathcal{Q}_0(0,\zeta)=\frac1{(1-\zeta)^{1-\gamma_1}},\quad\gamma_1=\frac12+\frac1{\pi}\arctan\left(\frac\mu a\right),
\end{equation}
which yields the same result obtained in \cite{LeDWie-2009}. We are left to consider the cases with $\alpha\neq1$ and $\mu\neq0$. We study $0<\alpha<1$ and $1<\alpha<2$ separately.

\subsubsection{$0<\alpha<1$.}
We prove that in this case $\mathcal{Q}(0,\zeta)$ can be put in the form
\begin{equation}\label{eq:App_Q_0_alpha1}
	\mathcal{Q}_0(0,\zeta)=\frac{\rme^{-\mathcal{U}_\alpha(\zeta;\mu,\beta)}}{(1-\zeta)^{1-\gamma}},\quad\gamma=\frac12+\frac1{\pi\alpha}\arctan[\beta\varpi(\alpha)],
\end{equation}
where $\mathcal{U}_\alpha(\zeta;\mu,\beta)$ is the power series
\begin{equation}
	\mathcal{U}_\alpha(\zeta;\mu,\beta)=\frac{1}{\pi}\sum_{n=1}^{\infty}\frac{\zeta^n}{n}U_\alpha(n;\mu,\beta),
\end{equation}
with $U_\alpha(n;\mu,\beta)$ denoting the integrals
\begin{eqnarray}
	U_\alpha(n;\mu,\beta)=\int_{0}^{\infty}\frac{dt}{t}\rme^{-t^\alpha}\Bigg\{&\sin\left[\beta\varpi(\alpha)t^\alpha+\frac{\mu t}{an^{1/\alpha-1}}\right]\nonumber\\
	&-\sin[\beta\varpi(\alpha)t^\alpha]\Bigg\}.
\end{eqnarray}
By analysing the regime $\zeta\to1$, one extracts the singular behaviour
\begin{equation}\label{eq:App_Q_0_alpha1_b}
	\mathcal{Q}_0(0,\zeta)\sim\frac{\rme^{-\mathcal{U}_\alpha(1;\mu,\beta)}}{(1-\zeta)^{1-\gamma}},
\end{equation}
and by an application of Tauberian theorems one deduces the large-$n$ decay
\begin{equation}\label{eq:App_q0_LS_a<1}
	q_0(n)\sim \frac{\rme^{-\mathcal{U}_\alpha(1;\mu,\beta)}}{\Gamma(1-\gamma)}n^{-\gamma}.
\end{equation}

To prove these results, we use \eref{eq:App_Omega_sum} to write explicitly
\begin{equation}
	-\rmi\Omega(0,\zeta)=-\frac1\pi\sum_{n=1}^{\infty}\frac{\zeta^n}{n}\int_{0}^{\infty}\frac{\rme^{-t^\alpha}}{t}\sin\left[\beta\varpi(\alpha)t^\alpha+\frac{\mu t}{an^{1/\alpha-1}}\right]dt.
\end{equation}
By summing and subtracting the integral of \eref{eq:App_Int_LS_mu0}, the integral at the rhs can be written as
\begin{eqnarray}
	\int_{0}^{\infty}\frac{\rme^{-t^\alpha}}{t}\sin\left[\beta\varpi(\alpha)t^\alpha+\frac{\mu t}{an^{1/\alpha-1}}\right]dt=&U_\alpha(n;\mu,\beta)\nonumber\\
	&-\frac1\alpha\arctan[\beta\varpi(\alpha)].
\end{eqnarray}
Thus,
\begin{equation}
	-\rmi\Omega(0,\zeta)=-\mathcal{U}_\alpha(\zeta;\mu,\beta)+\frac1{\pi\alpha}\arctan[\beta\varpi(\alpha)]\ln(1-\zeta),
\end{equation}
whence it follows \eref{eq:App_Q_0_alpha1}. We now show that $\mathcal{U}_\alpha(\zeta;\mu,\beta)$ tends to a finite limit for $\zeta\to1$. By using the identity \eref{eq:App_id_sum} to sum the series defining $\mathcal{U}_\alpha(\zeta;\mu,\beta)$, we have
 \begin{eqnarray}
 	\mathcal{U}_\alpha(\zeta;\mu,\beta)=&\frac1\pi\int_{0}^{\infty}\frac{dt}{t}\Bigg\{\arctan\left[\frac{\zeta\rme^{-t^\alpha}\sin\left(\mu t/a+\beta\varpi(\alpha) t^\alpha\right)}{1-\zeta\rme^{-t^\alpha}\cos\left(\mu t/a+\beta\varpi(\alpha)t^\alpha\right)}\right]\nonumber\\
 	&\quad-\arctan\left[\frac{\zeta\rme^{-t^\alpha}\sin\left(\beta\varpi(\alpha) t^\alpha\right)}{1-\zeta\rme^{-t^\alpha}\cos\left(\beta\varpi(\alpha)t^\alpha\right)}\right]\Bigg\}.
 \end{eqnarray}
 One can use this expression to show that, for $\zeta=1$, the integrand has an integrable singularity $\propto t^{-\alpha}$ at $t=0$, thus $\mathcal{U}_\alpha(1;\mu,\beta)$ is finite.  It follows that the singular behaviour of $\mathcal{Q}_0(0,\zeta)$ as $\zeta\to1$ is given by \eref{eq:App_Q_0_alpha1_b}, whence we deduce the large-$n$ decay of \eref{eq:App_q0_LS_a<1}.
 
 \subsubsection{$1<\alpha<2$.}\label{s:App_LS_a>1}
 \begin{figure}
 	\centering
 	\includegraphics[scale=0.7]{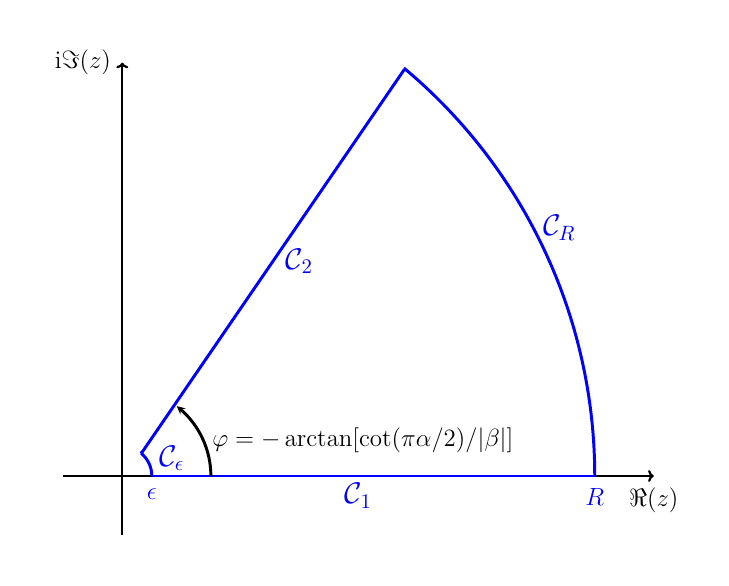}
 	\caption{Integration contour for the evaluation of the integral in \eref{eq:App_Contour2}. The angle of both arcs is $\varphi=-\arctan[\cot(\pi\alpha/2)/|\beta|]$, ensuring that the contribution on $\mathcal{C}_R$ vanishes in the limit $R\to\infty$.}
 	\label{fig:App_Contour2}
 \end{figure}
 Let us define
 \begin{equation}\label{eq:App_varphi}
 	 	\varphi\coloneq-\arctan\left[\frac1{|\beta|\varpi(\alpha)}\right]=-\arctan\left[\frac{1}{|\beta|}\cot\left(\frac{\pi\alpha}{2}\right)\right]\in\left(0,\frac\pi2\right).
 \end{equation}
Furthermore, let us introduce the parameters
 \begin{eqnarray}
 	A_\alpha(\mu,\beta)&\coloneq\sgn(\mu)\left(\frac{a}{|\mu|\cos\varphi}\right)^\alpha\frac{2\beta\varpi(\alpha)\Theta[-\sgn(\mu)\beta]}{\sqrt{1+\beta^2\varpi(\alpha)^2}}\\
 	B_\alpha(\mu,\beta)&\coloneq\left(\frac{a}{|\mu|\cos\varphi}\right)^\alpha\frac{1+\frac{\sgn(\beta)}{\sgn(\mu)}\beta^2\varpi(\alpha)^2}{\sqrt{1+\beta^2\varpi(\alpha)^2}},
 \end{eqnarray}
 where, in the first line, $\Theta(x)$ denotes the Theta function: $\Theta(x)=1$ for $x>0$ and $\Theta(x)=0$ otherwise. We prove that, depending on the sign of $\mu$, the generating function $\mathcal{Q}_0(0,\zeta)$ can be written as
 \begin{equation}\label{eq:App_Q0_a>1}
 	\mathcal{Q}_0(0,\zeta)=\cases{
 		\frac{\rme^{-\mathcal{V}_\alpha(\zeta;\mu,\beta)}}{1-\zeta} & $\mu<0$\\
 		\rme^{-\mathcal{V}_\alpha(\zeta;\mu,\beta)} & $\mu>0$
 	}
 \end{equation}
 where $\mathcal{V}_\alpha(\zeta;\mu,\beta)$ is
 \begin{equation}\label{eq:App_J_genfun_a>1}
 	\mathcal{V}_\alpha(\zeta;\mu,\beta)=\frac{\sgn(\mu)}\pi\sum_{n=1}^{\infty}\frac{\zeta^n}{n}V_\alpha(n;\mu,\beta),
 \end{equation}
 with $V_\alpha(n;\mu,\beta)$ denoting
 \begin{eqnarray}\label{eq:App_Jn_a>1}
 	V_\alpha(n;\mu,\beta)=\alpha\int_{0}^{\infty}\frac{dt}{t}&\rme^{-t\tan\left(\frac{\varphi}{\alpha}\right)}\times\nonumber\\
 	&\left\{\rme^{-\frac{A_\alpha(\mu,\beta)t^\alpha}{n^{\alpha-1}}}\sin\left[t-\frac{B_\alpha(\mu,\beta)t^\alpha}{n^{\alpha-1}}\right]-\sin(t)\right\}.
 \end{eqnarray}
 It follows that we have two distinct asymptotic behaviours for $q_0(n)$:
 \begin{enumerate}
 	\item For $\mu>0$, we find that $q_0(n)$ exhibits the power-law decay
 	\begin{equation}\label{eq:App_q0_a>1_mu>0}
 		q_0(n)\sim \left[A_\alpha(\mu,\beta)\mathcal{S}_\alpha(\beta)+B_\alpha(\mu,\beta)\mathcal{C}_\alpha(\beta)\right]\frac{\rme^{-\mathcal{V}_\alpha(1;\mu,\beta)}}{\pi n^{\alpha}},
 	\end{equation}
 	where $\mathcal{C}_\alpha(\beta)$ and $\mathcal{S}_\alpha(\beta)$ are the integrals defined by
 	\begin{eqnarray}
 		\mathcal{C}_\alpha(\beta)&=\alpha\int_{0}^{\infty}\frac{\rme^{-t\,\tan\left(\frac{\varphi}{\alpha}\right)}}{t^{1-\alpha}}\cos(t)dt\label{eq:App_C_integral}\\
 		\mathcal{S}_\alpha(\beta)&=\alpha\int_{0}^{\infty}\frac{\rme^{-t\,\tan\left(\frac{\varphi}{\alpha}\right)}}{t^{1-\alpha}}\sin(t)dt.\label{eq:App_S_integral}
 	\end{eqnarray}
 	\item For $\mu<0$, in the long time limit the survival probability converges to a constant, viz.
 	\begin{equation}\label{eq:App_q0_a>1_mu<0}
 		q_0(n)\to\exp\left[-\mathcal{V}_\alpha(1;\mu,\beta)\right].
 	\end{equation}
 \end{enumerate}
 
To prove these results, let us write \eref{eq:App_Omega_sum} as
 \begin{equation}\label{eq:App_Omega_sum_a>1}
 	-\rmi\Omega(0,\zeta)=-\frac{\sgn(\mu)}{\pi\alpha}\sum_{n=1}^{\infty}\frac{\zeta^n}{n}\int_{0}^{\infty}\frac{\rme^{-c_1t/n^{\alpha-1}}}{t}\sin\left(t^{1/\alpha}+\frac{c_2t}{n^{\alpha-1}}\right)dt,
 \end{equation}
 where, for simplicity, we set
 \begin{eqnarray}
 	c_1 &=\left(\frac{a}{|\mu|}\right)^\alpha\label{eq:App_c}\\
 	c_2 &=\frac{\beta\varpi(\alpha)}{\sgn(\mu)}c_1.\label{eq:App_C}
 \end{eqnarray}
 To extract the large-$n$ behaviour of the integral at the rhs of \eref{eq:App_Omega_sum_a>1}, we consider the contour integral
 \begin{equation}\label{eq:App_Contour2}
 	\int_{\mathcal{C}}dz f(z)=\int_{\mathcal{C}}\frac{dz}{z}\exp\left(-\frac{c_1z}{n^{\alpha-1}}+\rmi z^{1/\alpha}+\frac{\rmi c_2z}{n^{\alpha-1}}\right),
 \end{equation}
 where $\mathcal{C}$ is the contour in \fref{fig:App_Contour2}, consisting of the arcs of two circumferences centred at the origin, of radius $\epsilon$ and $R$ respectively, both of angle $\varphi$, connected by two line segments $\mathcal{C}_1$ and $\mathcal{C}_2$. Note that the integral at the rhs of \eref{eq:App_Omega_sum_a>1} is
 \begin{equation}\label{eq:App_Int_vs_IntCont}
 	\int_{0}^{\infty}\frac{\rme^{-c_1t/n^{\alpha-1}}}{t}\sin\left(t^{1/\alpha}+\frac{c_2t}{n^{\alpha-1}}\right)dt=\lim_{R\to\infty,\,\epsilon\to0}\Im\left[\int_{\mathcal{C}_1}dzf(z)\right].
 \end{equation}
One can show that in the limit $R\to\infty$ the contribution on $\mathcal{C}_R$ vanishes, while
 \begin{equation}
 	\lim_{\epsilon\to0}\int_{\mathcal{C}_\epsilon}dzf(z)\to-\rmi\varphi.
 \end{equation}
 Therefore, by the residue theorem we have
 \begin{eqnarray}
 	\lim_{R\to\infty,\,\epsilon\to0}\int_{\mathcal{C}_1}dzf(z)&=-\lim_{\epsilon\to0}\int_{\mathcal{C}_\epsilon}dzf(z)-\lim_{R\to\infty,\epsilon\to0}\int_{\mathcal{C}_2}dzf(z),\label{eq:App_residueTHM}\\
 	&=\rmi\varphi+\int_{0}^{\infty}dt\rme^{\rmi\varphi}f\left(t\rme^{\rmi\varphi}\right)\\
 	&=\rmi\varphi+\alpha\int_{0}^{\infty}dt\frac{t^{\alpha-1}\rme^{\rmi\varphi}}{\cos^\alpha(\varphi)}f\left(\frac{t^\alpha\rme^{\rmi\varphi}}{\cos^\alpha(\varphi)}\right),
 \end{eqnarray}
 where the change of variable $t\to t^{\alpha}/\cos^\alpha(\varphi)$ in the third line has been made for future convenience. The imaginary part of the third line is written explicitly as
 \begin{equation}\label{eq:App_long1}
 	\varphi+\alpha\int_{0}^{\infty}\frac{dt}{t}\rme^{-t\tan\left(\frac{\varphi}{\alpha}\right)-\frac{(c_1\cos\varphi+c_2\sin\varphi)t^\alpha}{\cos^\alpha(\varphi)n^{\alpha-1}}}\sin\left[t-\frac{\left(c_1\sin\varphi-c_2\cos\varphi\right)t^{\alpha}}{\cos^\alpha(\varphi)n^{\alpha-1}}\right],
 \end{equation}
and furthermore, by combining \eref{eq:App_varphi}, \eref{eq:App_c} and \eref{eq:App_C}, we have the relations
 \begin{eqnarray}
 	\frac{c_1\cos\varphi+c_2\sin\varphi}{\cos^\alpha(\varphi)}&=A_\alpha(\mu,\beta)\\
 	\frac{c_1\cos\varphi-c_2\sin\varphi}{\cos^\alpha(\varphi)}&=B_\alpha(\mu,\beta).
 \end{eqnarray}
Now, the integral in \eref{eq:App_long1} converges for $n\to\infty$ to the integral
\begin{equation}
	\alpha\int_{0}^{\infty}\frac{dt}{t}\rme^{-t\tan\left(\frac{\varphi}{\alpha}\right)}\sin(t)=\alpha\left(\frac{\pi}{2}-\frac{\varphi}{\alpha}\right),
\end{equation}
thus \eref{eq:App_long1} can be rewritten as $\case{\pi\alpha}2+V_\alpha(n;\mu,\beta)$, see \eref{eq:App_Jn_a>1}. Therefore, \eref{eq:App_Int_vs_IntCont} yields
\begin{equation}
	\int_{0}^{\infty}\frac{\rme^{-c_1t/n^{\alpha-1}}}{t}\sin\left(t^{1/\alpha}+\frac{c_2t}{n^{\alpha-1}}\right)dt=\frac{\pi\alpha}{2}+V_\alpha(n;\mu,\beta).
\end{equation}
 By going back to \eref{eq:App_Omega_sum_a>1}, it follows that we can rewrite $-\rmi\Omega(0,\zeta)$ as
\begin{equation}
	-\rmi\Omega(0,\zeta)=\frac{\sgn(\mu)}{2}\ln(1-\zeta)-\mathcal{V}_\alpha(\zeta;\mu,\beta),
\end{equation}
hence we obtain the expression of $\mathcal{Q}_0(0,\zeta)$ given in \eref{eq:App_Q0_a>1}. We now show that $\mathcal{V}_\alpha(1;\mu,\beta)$ is convergent. By expanding the term in brackets of \eref{eq:App_Jn_a>1} for large $n$, we get
\begin{equation}\label{eq:App_Jn_a>1_asymp}
		V_\alpha(n;\mu,\beta)\sim-K_\alpha(\mu,\beta)n^{1-\alpha}+o\left(n^{1-\alpha}\right),
\end{equation}
with
\begin{equation}
	K_\alpha(\mu,\beta)=A_\alpha(\mu,\beta)\mathcal{S}_\alpha(\beta)+B_\alpha(\mu,\beta)\mathcal{C}_\alpha(\beta),
\end{equation}
where $\mathcal{C}_\alpha(\beta)$ and $\mathcal{S}_\alpha(\beta)$ are the integrals defined by \eref{eq:App_C_integral} and \eref{eq:App_S_integral}. Thus, the coefficients of the power series defining $\mathcal{V}_\alpha(n;\mu,\beta)$ are $\propto n^{-\alpha}$ for large $n$, thus $\mathcal{V}_\alpha(1;\mu,\beta)$ is convergent. Finally, we prove \eref{eq:App_q0_a>1_mu>0} and \eref{eq:App_q0_a>1_mu<0}.  For $\mu>0$, we have
\begin{equation}
	\mathcal{Q}_0(0,\zeta)\sim\rme^{-\mathcal{V}_\alpha(1;\mu,\beta)},
\end{equation}
therefore $\mathcal{Q}_0(0,\zeta)$ converges to a constant. To use Tauberian theorems, we need to consider a divergent series as $\zeta\to1$, thus we take the derivative with respect to $\zeta$:
\begin{eqnarray}
	\mathcal{Q}'_0(0,\zeta)=\sum_{n=1}^{\infty}nq_0(n)\zeta^{n-1}&=\rme^{-\mathcal{V}_\alpha(\zeta;\mu,\beta)}\left[-\frac{\partial\mathcal{V}_\alpha(\zeta;\mu,\beta)}{\partial\zeta}\right]\\
	&=\rme^{-\mathcal{V}_\alpha(\zeta;\mu,\beta)}\left[-\frac1{\pi}\sum_{n=1}^{\infty}\zeta^{n-1}V_\alpha(n;\mu,\beta)\right].
\end{eqnarray}
Given the asymptotic behaviour of $V_\alpha(n;\mu,\beta)$ in \eref{eq:App_Jn_a>1_asymp}, it follows that the large-$n$ behaviour of the coefficients is
\begin{equation}
	nq_0(n)\sim\frac{\rme^{-\mathcal{V}_\alpha(1;\mu,\beta)}}{\pi}K_\alpha(\mu,\beta)n^{1-\alpha},
\end{equation}
whence it follows \eref{eq:App_q0_a>1_mu>0}. For $\mu<0$ instead, the generating function $\mathcal{Q}_0(0,\zeta)$ when $\zeta\to1$ behaves as
\begin{equation}
	\mathcal{Q}_0(0,\zeta)\sim\frac{\rme^{-\mathcal{V}_\alpha(1;\mu,\beta)}}{1-\zeta},
\end{equation}
where the numerator is a constant, thus \eref{eq:App_q0_a>1_mu<0} follows immediately.

\subsection{General Lévy flights}
We now extend the results to general Lévy flights. We recall that we consider characteristic functions of the form $\FT{\wp}(k)=\wpr(k)+\rmi\wpi(k)$, with $\wpr(k)\sim1-|ak|^\alpha$ and $\wpi(k)$ given by \eref{eq:CF_Im_small-k} for small $k$. Let us remark that, by writing $\FT{\wp}(k)$ in the form $\FT{\wp}(k)=\rho(k)\rme^{\rmi\vartheta(k)}$, we have for small $k$:
\begin{equation}
	\rho(k)\sim1-|ak|^\alpha,\quad\vartheta(k)\sim\cases{\beta\varpi(\alpha)\sgn(k)|ak|^\alpha & $0<\alpha<1$\\
		\mu k & $\alpha=1$\\
		\mu k+\beta\varpi(\alpha)\sgn(k)|ak|^\alpha&$1<\alpha<2$}
\end{equation}

\subsubsection{$0<\alpha<1$.}
Let us consider the integral at the rhs of \eref{eq:App_Omega_sum}. With the change of integration variable $t\to t/(an)^{1/\alpha}$, in the large-$n$ limit we have
\begin{eqnarray}
	\fl	\int_{0}^{\infty}\frac{dt}{t}\rho\left(\frac{t}{(an)^{1/\alpha}}\right)^n\sin\left[n\vartheta\left(\frac{t}{(an)^{1/\alpha}}\right)\right]&\sim\int_{0}^{\infty}\frac{dt}{t}\rme^{-t^\alpha}\sin\left[\beta\varpi(\alpha)t^\alpha\right]\\
	&\sim\frac1\alpha\arctan[\beta\varpi(\alpha)],
\end{eqnarray}
thus, for $\zeta\to1$
\begin{equation}
	\rmi\Omega(0,\zeta)\sim-\frac1{\pi\alpha}\arctan[\beta\varpi(\alpha)]\ln(1-\zeta).
\end{equation}
Therefore, we multiply and divide $\mathcal{Q}_0(0,\zeta)$ by $\rme^{\frac1{\pi\alpha}\arctan[\beta\varpi(\alpha)]\ln(1-\zeta)}$, obtaining
\begin{equation}
	\mathcal{Q}_0(0,\zeta)=\frac{\rme^{-\mathcal{U}_\alpha^*(\zeta;\mu,\beta)}}{(1-\zeta)^{1-\gamma}},\quad\gamma=\frac12+\frac1{\pi\alpha}\arctan[\beta\varpi(\alpha)],
\end{equation}
where $\mathcal{U}_\alpha^*(\zeta;\mu,\beta)$ is the series:
\begin{equation}
	\mathcal{U}_\alpha^*(\zeta;\mu,\beta)=\frac{1}{\pi}\sum_{n=1}^{\infty}\frac{\zeta^n}{n}\left\{\int_{0}^{\infty}\frac{\rho(t)^n}{t}\sin[n\vartheta(t)]dt-\frac1\alpha\arctan[\beta\varpi(\alpha)]\right\}.
\end{equation}
By employing the identity
\begin{equation}
	\frac1{\alpha}\arctan(\vartheta)\ln(1-\zeta)=-\sum_{n=1}^{\infty}\frac{\zeta^n}{n}\int_{0}^{\infty}\frac{dt}{t}\rme^{-nt^\alpha}\sin(n\vartheta t^{\alpha}),
\end{equation}
and then summing the series by using \eref{eq:App_id_sum}, we rewrite $\mathcal{U}_\alpha^*(\zeta;\mu,\beta)$ as the integral
\begin{eqnarray}
	\mathcal{U}_\alpha^*(\zeta;\mu,\beta)=\frac1{\pi}\int_{0}^{\infty}\frac{dt}{t}\Bigg\{&\arctan\left[\frac{\zeta\wpr(t)}{1-\zeta\wpi(t)}\right]\nonumber\\
	&-\arctan\left[\frac{\zeta\rme^{-t^\alpha}\sin(\beta\varpi(\alpha)t^\alpha)}{1-\zeta\rme^{-t^\alpha}\cos(\beta\varpi(\alpha)t^\alpha)}\right]\Bigg\}.
\end{eqnarray}
When $\zeta\to1$, the integrand has an integrable singularity at $t=0$, thus $\mathcal{U}_\alpha^*(1;\mu,\beta)$ is finite.  Hence, we conclude that in the limit $\zeta\to1$ the generating function behaves as
\begin{equation}
	\mathcal{Q}_0(0,\zeta)\sim\frac{\rme^{-\mathcal{U}_\alpha^*(1;\mu,\beta)}}{(1-\zeta)^{1-\gamma}},
\end{equation}
whence we deduce
\begin{equation}
	q_0(n)\sim\frac{\rme^{-\mathcal{U}_\alpha^*(1;\mu,\beta)}}{\Gamma(1-\gamma)}n^{-\gamma}.
\end{equation}

\subsubsection{$\alpha=1$.} With the change of integration variable $t\to t/(an)$ in the integral at the rhs of \eref{eq:App_Omega_sum}, in the large-$n$ limit we have
\begin{eqnarray}
	\int_{0}^{\infty}\frac{dt}{t}\rho\left(\frac{t}{an}\right)^n\sin\left[n\vartheta\left(\frac{t}{an}\right)\right]&\sim\int_{0}^{\infty}\frac{dt}{t}\rme^{-t}\sin\left(\frac{\mu t}a\right)\\
	&\sim\arctan\left(\frac\mu a\right),
\end{eqnarray}
and thus, following the steps of the previous case, we arrive at
\begin{equation}
	\mathcal{Q}_0(0,\zeta)=\frac{\rme^{-\mathcal{W}(\zeta;\mu)}}{(1-\zeta)^{1-\gamma_1}},\quad\gamma_1=\frac12+\frac1{\pi}\arctan\left(\frac\mu a\right),
\end{equation}
where $\mathcal{W}(\zeta;\mu)$ is
\begin{equation}
	\mathcal{W}(\zeta;\mu)=\frac{1}{\pi}\sum_{n=1}^{\infty}\frac{\zeta^n}{n}\left\{\int_{0}^{\infty}\frac{\rho(t)^n}{t}\sin[n\vartheta(t)]dt-\arctan\left(\frac\mu a\right)\right\}.
\end{equation}
Again, one can rewrite $\mathcal{W}(\zeta;\mu)$ as an integral and show that for $\zeta\to1$ the resulting integral is convergent. Therefore, we finally obtain
\begin{equation}
	q_0(n)\sim\frac{\rme^{-\mathcal{W}(1;\mu)}}{\Gamma(1-\gamma_1)}n^{-\gamma_1}.
\end{equation}

\subsubsection{$1<\alpha<2$.} Now we employ the change of variable $t\to t^{1/\alpha}/(an)$ in the integral at the rhs of \eref{eq:App_Omega_sum}. In the large-$n$ limit,
\begin{eqnarray}
	\fl	\int_{0}^{\infty}\frac{dt}{t}\rho\left(\frac{t^{1/\alpha}}{an}\right)^n\sin\left[n\vartheta\left(\frac{t^{1/\alpha}}{an}\right)\right]&\sim\frac{\sgn(\mu)}{\alpha}\int_{0}^{\infty}\frac{dt}{t}\rme^{-c_1t/n^{\alpha-1}}\sin\left(t^{1/\alpha}+\frac{c_2t}{n^{\alpha-1}}\right)\\
	&\sim\frac{\sgn(\mu)}{\alpha}\left[\frac{\pi\alpha}{2}+V_\alpha(n;\mu,\beta)\right],\label{eq:Approx_Int_GenLévy_a>1}
\end{eqnarray}
where, in the first line, the integral at the rhs corresponds to the integral in \eref{eq:App_Omega_sum_a>1}, and $c_1$ and $c_2$ are the constants given in \eref{eq:App_c} and \eref{eq:App_C}, respectively. To pass from the first to the second line, we followed the discussion of \ref{s:App_LS_a>1}. By plugging the second line into \eref{eq:App_Omega_sum} and recalling that $V_\alpha(n;\mu,\beta)\propto n^{1-\alpha}$, we can say that the singular behaviour of $-\rmi\Omega(0,\zeta)$ as $\zeta\to1$ is
\begin{equation}
	-\rmi\Omega(0,\zeta)\sim\frac{\sgn(\mu)}2\ln(1-\zeta).
\end{equation}
Thus, we rewrite
\begin{equation}\label{eq:App_Q0_a>1_general}
	\mathcal{Q}_0(0,\zeta)=\cases{
		\frac{\rme^{-\mathcal{V}_\alpha^*(\zeta;\mu,\beta)}}{1-\zeta} & $\mu<0$\\
		\rme^{-\mathcal{V}_\alpha^*(\zeta;\mu,\beta)} & $\mu>0$
	}
\end{equation}
where
\begin{equation}
	\mathcal{V}_\alpha^*(\zeta;\mu,\beta)=\sum_{n=1}^{\infty}\frac{\zeta^n}{n}\left\{\frac1{\pi}\int_{0}^{\infty}\frac{dt}{t}\rho(t)^n\sin[n\vartheta(t)]-\frac{\sgn(\mu)}2\right\},
\end{equation}
which is convergent for $\zeta=1$. It follows that:
\begin{enumerate}
	\item for $\mu>0$, let us take the derivative with respect to $\zeta$:
	\begin{equation}
		\mathcal{Q}'_0(0,\zeta)=-\rme^{-\mathcal{V}_\alpha^*(\zeta;\mu,\beta)}\sum_{n=1}^{\infty}\zeta^{n-1}\left\{\frac1{\pi}\int_{0}^{\infty}\frac{dt}{t}\rho(t)^n\sin[n\vartheta(t)]-\frac{1}2\right\}.
	\end{equation}
	According to \eref{eq:Approx_Int_GenLévy_a>1}, for large $n$ the term in brackets has a leading behaviour given by $V_\alpha(n;\mu,\beta)/(\pi\alpha)$. Therefore, $q_0(n)$ displays the asymptotic power-law decay:
	\begin{equation}
		q_0(n)\sim K_\alpha(\mu,\beta)\frac{\rme^{-\mathcal{V}_\alpha^*(1;\mu,\beta)}}{\pi n^\alpha}.
	\end{equation}
	\item for $\mu<0$, the survival probability converges to the constant, given by:
	\begin{equation}
		q_0(n)\to\exp[-\mathcal{V}_\alpha^*(1;\mu,\beta)].
	\end{equation}
\end{enumerate}
 
\section{Leap-over distribution for symmetric Lévy flights}\label{s:App_Leap_LF}
\subsection{Expansion of the characteristic function}
Here we prove \eref{eq:Leap_PDF_Lévy_FT}. We recall that $\FT{\mathcal{L}}_b(k)=\rme^{-\rmi kb}\mathcal{F}_b(k,1)$, where
\begin{equation}\label{eq:App_Fb_cont}
	\mathcal{F}_b(k,1)=1-\sqrt{1-\FT{\wp}(k)}\rme^{-\rmi\Omega(k,1)}\mathcal{H}_b(k,1).
\end{equation}
We first determine the small-$k$ expansion of the phase, which for symmetric jumps can be written as
\begin{equation}
	\Omega(k,1)=-\frac{k}{\pi}\int_{0}^{\infty}\frac{\ln[1-\FT{\wp}(t)]}{k^2-t^2}dt=-\frac{\sgn(k)}{\pi}\int_{0}^{\infty}\frac{\ln[1-\FT{\wp}(|k|t)]}{1-t^2}dt.
\end{equation}
We first note that
\begin{equation}
	\frac1\pi\int_{0}^{\infty}\frac{\ln(|ak|^\alpha t^\alpha)}{1-t^2}dt=-\frac{\pi\alpha}{4},
\end{equation}
because the principal value of the integral of $1/(1-t^2)$ vanishes, while
\begin{equation}
	\int_{0}^{\infty}\frac{\ln(t)}{1-t^2}dt=2\int_{0}^{1}\frac{\ln(t)}{1-t^2}dt=-\frac{\pi^2}{4},
\end{equation}
see \textbf{4.231}(13) in \cite{GraRyz}. Therefore,
\begin{equation}\label{eq:App_Leap_fund}
	\Omega(k,1)=\sgn(k)\frac{\pi\alpha}{4}-\sgn(k)\frac{1}{\pi}\int_{0}^{\infty}\ln\left[\frac{1-\FT{\wp}(|k|t)}{|ak|^\alpha t^\alpha}\right]\frac{dt}{1-t^2}.
\end{equation}
We now show that in the small-$k$ limit, the second term yields a vanishing contribution. Let us suppose that for small $k$:
\begin{equation}\label{eq:App_Leap_CFapprox}
	\frac{1-\FT{\wp}(k)}{|ak|^\alpha}\sim1-C|k|^\nu,\quad0<\nu<2.
\end{equation}
If $0<\nu\leq1$, in the second integral we split the integration domain in $(0,1)$, $(1,1/|k|)$ and $(1/|k|,\infty)$. The integral in the last domain is equal to
\begin{equation}
	-\frac k\pi\int_{1}^{\infty}\ln\left[\frac{1-\FT{\wp}(t)}{a^\alpha t^\alpha}\right]\frac{dt}{k^2-t^2},
\end{equation}
which yields a contribution of order $k$ in the small-$k$ limit. In the first two domains, we can use \eref{eq:App_Leap_CFapprox} to approximate
\begin{eqnarray}
\fl	\int_{0}^{1^-}+\int_{1^+}^{1/|k|}\ln\left[\frac{1-\FT{\wp}(|k|t)}{|ak|^\alpha t^\alpha}\right]\frac{dt}{1-t^2}&\approx\int_{0}^{1^-}+\int_{1^+}^{1/|k|}\frac{\ln(1-C|k|^\nu t^\nu)}{1-t^2}dt\\
	&\approx-C|k|^\nu\int_{0}^{1^-}+\int_{1^+}^{1/|k|}\frac{t^\nu}{1-t^2}dt.
\end{eqnarray}
For $\nu<1$, in the second integral at the second line we perform the change of variable $t\to1/t$ and take the limit $k\to0$, obtaining
\begin{eqnarray}
	-C|k|^\nu\int_{0}^{1^-}+\int_{1^+}^{1/|k|}\frac{t^\nu}{1-t^2}dt&\sim -C|k|^{\nu}\int_{0}^{1}\frac{t^{\nu}-t^{-\nu}}{1-t^2}dt\\
	&\sim\frac{\pi}{2}\tan\left(\frac{\pi\nu}{2}\right)C|k|^\nu,
\end{eqnarray}
where the rhs of the first and second line are equal, see integral \textbf{3.244}(2) in \cite{GraRyz}. For $\nu=1$ we have
\begin{equation}
	\int_{0}^{1^-}+\int_{1^+}^{1/|k|}\frac{t}{1-t^2}dt=\frac12\ln\left(\frac{k^2}{1-k^2}\right),
\end{equation}
thus for small $k$:
\begin{equation}
	-C|k|\int_{0}^{1^-}+\int_{1^+}^{1/|k|}\frac{t}{1-t^2}dt\sim -C|k|\ln(|k|).
\end{equation}
If $1<\nu<2$, we rewrite the integral at the rhs of \eref{eq:App_Leap_fund} by changing the integration variable $t\to t/|k|$, and then note that for $k\to0$
\begin{equation}
	\frac{1}{\pi}\int_{0}^{\infty}\ln\left[\frac{1-\FT{\wp}(t)}{a^\alpha t^\alpha}\right]\frac{dt}{k^2-t^2}\to \Lambda\coloneq-\frac{1}{\pi}\int_{0}^{\infty}\ln\left[\frac{1-\FT{\wp}(t)}{a^\alpha t^\alpha}\right]\frac{dt}{t^2}.
\end{equation}
Thus, for small $k$:
\begin{equation}
	\sgn(k)\frac{1}{\pi}\int_{0}^{\infty}\ln\left[\frac{1-\FT{\wp}(|k|t)}{|ak|^\alpha t^\alpha}\right]\frac{dt}{1-t^2}\sim \Lambda k.
\end{equation}
Putting everything together, we have that for small $k$:
\begin{equation}\label{eq:App_Omega_small-k}
	\Omega(k,1)\sim\sgn(k)\frac{\pi\alpha}{2}-\cases{
	\frac{C}{2}\tan\left(\frac{\pi\nu}{2}\right)\sgn(k)|k|^{\nu} & $0<\nu<1$\\
	-\frac{C}{\pi}k\ln(|k|) & $\nu=1$\\
	\Lambda k & $1<\nu<2$
	}
\end{equation}
Thus,
\begin{equation}
	\rme^{-\rmi\Omega(k,1)}\sim\left[\cos\left(\frac{\pi\alpha}{4}\right)-\rmi\sgn(k)\sin\left(\frac{\pi\alpha}{4}\right)\right]\rme^{\rmi\omega(k)},
\end{equation}
where $\omega(k)$ vanishes for $k\to0$. Expanding \eref{eq:App_Fb_cont} and noting that $\mathcal{H}_b(k,1)$ contributes, at lowest order and for fixed $b$, only with the zero-order term $\mathcal{H}_b(0,1)$, we obtain the expansion \eref{eq:Leap_PDF_Lévy_FT}.

\subsection{Approximate expressions for $\mathcal{H}_b(0,1)$}
According to \eref{eq:Hb}, the small-$b$ behaviour of $\mathcal{H}_b(0,1)$ can be obtained from the large-$u$ behaviour of $1/\Psi_0^+(\rmi u,1)$, whose expression, for symmetric jump distributions, can be simplified to
\begin{equation}
	\frac1{\Psi_0^+(\rmi u,1)}=\exp\left\{-\frac u\pi\int_{0}^{\infty}\frac{\ln[1-\FT{\wp}(t)]}{t^2+u^2}dt\right\}.
\end{equation}
Rescaling $u\to u/a$, for large $u/a$ we have
\begin{equation}
	\frac1{\Psi_0^+(\rmi u/a,1)}\approx\exp\left\{-\frac a{\pi u}\int_{0}^{\infty}\ln[1-\FT{\wp}(t)]dt\right\}=\exp\left(\frac{\lambda}{u}\right),
\end{equation}
where $\lambda$ is defined by \eref{eq:Leap_lambda}. Therefore, for small $b$ we can approximate
\begin{equation}
	\mathcal{H}_b(0,1)\approx\frac1{2\pi\rmi}\int_{\mathcal{B}}\frac{\rme^{ub/a+\lambda/u}}{u}du=I_0\left(2\sqrt{\lambda b/a}\right),
\end{equation}
where the inverse Laplace transformation has been performed according to formula $(14)$, pag. $197$ in \cite{Bat-I}.

Conversely, the large-$b$ behaviour of $\mathcal{H}_b(0,1)$ can be obtained from the small-$u$ behaviour of $1/\Psi_0^+(\rmi u,1)$. In this case
\begin{eqnarray}
\fl	\frac1{\Psi_0^+(\rmi u,1)}=\exp\left\{-\frac 1{\pi }\int_{0}^{\infty}\frac{\ln[1-\FT{\wp}(ut)]}{t^2+1}dt\right\}&\approx\exp\left[-\frac \alpha{\pi }\int_{0}^{\infty}\frac{\ln(aut)}{t^2+1}dt\right]\\
&\approx\exp\left[-\frac {\alpha}{2}\ln(au)\right],
\end{eqnarray}
where at the second line we used
\begin{equation}
	\int_{0}^{\infty}\frac{\ln(t)}{t^2+1}dt=0.
\end{equation}
Hence for large $b$
\begin{equation}
	\mathcal{H}_b(0,1)\approx\frac1{2\pi\rmi a^{\alpha/2}}\int_{\mathcal{B}}\frac{\rme^{ub}}{u^{1+\alpha/2}}du=\frac{(b/a)^{\alpha/2}}{\Gamma(1+\alpha/2)}.
\end{equation}

\section*{References}
\bibliographystyle{iopart-num}
\bibliography{Biblio.bib}
\end{document}